\documentclass{article}
\usepackage{cite} 
\usepackage[a4paper, margin=1.5cm]{geometry} 
\usepackage{graphicx} 
\usepackage{amsmath, amssymb} 
\usepackage{hyperref} 
\usepackage{xcolor} 
\usepackage{abstract} 
\usepackage{titlesec} 
\usepackage{float}
\usepackage{authblk}
\usepackage{subcaption} 
\usepackage{listings}
\titleformat{\section}{\bfseries\LARGE}{\thesection}{1em}{} 
\titleformat{\subsection}{\bfseries\Large}{\thesubsection}{1em}{} 

\title{ReadMe.LLM: A Framework to Help LLMs Understand Your Library}

\author{
    Sandya Wijaya \quad Jacob Bolano \quad Alejandro Gomez Soteres \\
    Shriyanshu Kode \quad Yue Huang \quad Anant Sahai
}

\affil{University of California, Berkeley}

\date{}

\begin{document}

\maketitle
\begin{abstract}
Large Language Models (LLMs) often struggle with code generation tasks involving niche software libraries. Existing code generation techniques with only human-oriented documentation can fail -- even when the LLM has access to web search and the library is documented online. To address this challenge, we propose ReadMe.LLM, LLM-oriented documentation for software libraries. By attaching the contents of ReadMe.LLM to a query, performance consistently improves to near-perfect accuracy, with one case study demonstrating up to 100\% success across all tested models. We propose a software development lifecycle where LLM-specific documentation is maintained alongside traditional software updates. In this study, we present two practical applications of the ReadMe.LLM idea with diverse software libraries, highlighting that our proposed approach could generalize across programming domains.  
\end{abstract}

\vspace{1em}

\section{Introduction}
Large Language Models (LLMs) like GPT \cite{10.5555/3495724.3495883} and Llama \cite{touvron2023llamaopenefficientfoundation} have transformed the software development ecosystem. More engineers are using LLMs to generate code with existing software libraries, leveraging these tools to approach coding tasks more efficiently and intuitively. In some cases, we are even seeing AI agents begin to replace human developers themselves. 
These models, often used as coding assistants, are capable of generating code, debugging, and creating documentation through natural language prompting. 

These advances build on a growing lineage of code-specific language models. OpenAI’s Codex \cite{chen2021evaluating}, fine-tuned on GitHub code, enabled natural language-to-code translation and powers tools like GitHub Copilot \cite{githubcopilot}. Meta’s Code LLaMA \cite{roziere2023code} was pretrained specifically on code-related data, allowing it to support multiple programming languages and longer context windows. Both models expanded the capabilities of LLMs while maintaining low-friction interfaces for developers — but they also highlight a core challenge: for this innovation and performance to continue, LLMs rely on consistent and structured documentation. Recent work has emphasized that structured specification (precise descriptions of a component’s expected behavior, inputs, and outputs) is essential to making LLM-based systems more modular and reliable \cite{stoica2024specification}. As these systems become increasingly integrated into IDEs and developer workflows, new coding practices continue to emerge.

One emerging trend, frequently referred to as ``vibe coding” \cite{Roose2025}, involves engineers prompting LLMs with simple, high-level natural language instructions and iteratively refining their code based on the model’s suggestions. This interactive exploratory approach enables fast prototyping and creates a more fluid software development process. 
\subsection{Challenge}
However, not all libraries are equally represented in LLM training data. Well-established libraries like Pandas \cite{mckinney2012python} have plenty of public documentation, Stack Overflow questions, and other resources that are ingested during LLM pretraining, allowing the LLM to produce reliable output, while lesser-known libraries are often misused or misrepresented in AI-generated code \cite{prompthub2024llms, latendresse2024chatgptgoodsoftwarelibrarian, abbassi2025unveilinginefficienciesllmgeneratedcode}. This gap negatively impacts both engineers and library developers. Engineers receive incorrect code, leading to frustration, prolonged debugging, and increased company resource expenditure \cite{10.1145/3597503.3608128}. Meanwhile, library developers risk losing potential users who abandon their tools in favor of alternatives that work seamlessly with LLM-generated code. 

In addition, as AI agents and services become more popular and increasingly integrated into development, their reliance on LLMs amplifies the underrepresentation of smaller libraries. If these agents struggle with less-documented tools, workflows become inefficient, reinforcing a cycle where only well-known libraries thrive. 

This dynamic is reshaping the entire software ecosystem. Smaller libraries lose potential users not due to their technical merit but because LLMs fail to capture them accurately. For engineers, this means fewer viable options and slower innovation. Our work addresses these systemic consequences by creating a framework that ensures LLMs can correctly understand and utilize any software library, leveling the playing field and fostering a more accessible development landscape. 
\subsection{Existing Solutions}
\begin{figure}[H]
    \centering
    \includegraphics[width=0.45\textwidth]{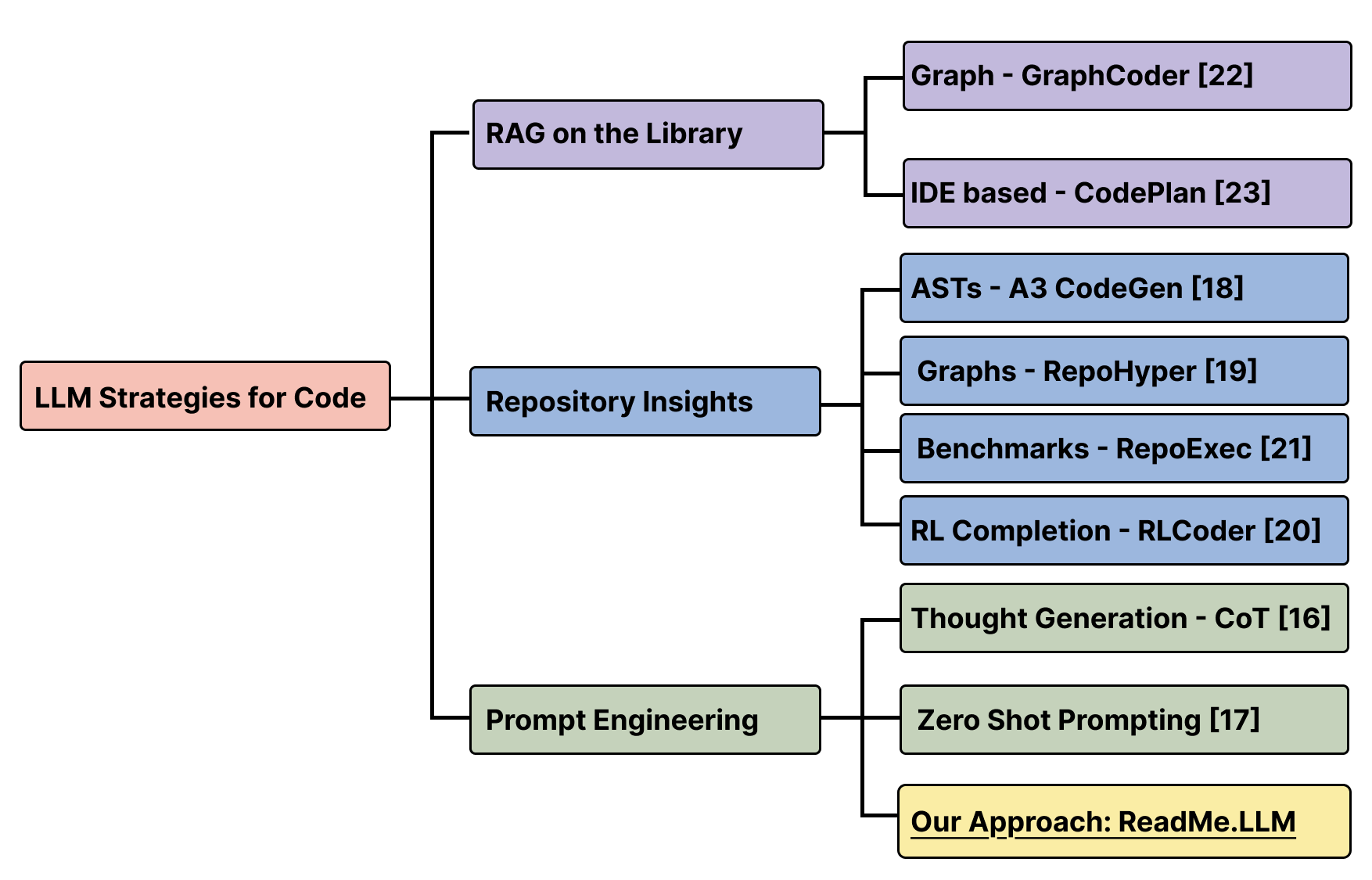}
    \caption{Survey of existing prompting strategies for code generation}
\end{figure}
There are different techniques for leveraging LLMs for code generation tasks. We list key categories and examples in Figure 1. Many strategies, such as Retrieval Augmented Generation (RAG) \cite{10.5555/3495724.3496517}, require additional infrastructure that falls outside of the typical developer workflow using standard IDEs\footnote{Of course, one possible view of the future would involve building a well standardized approach to integrating RAG with IDEs and copilots \cite{wang2025code}. But we are not there yet.}. The most accessible approach that offers the least friction to development is Prompt Engineering \cite{chen2025promptware}. Within Prompt Engineering, there has been research on thought generation, such as Chain of Thought \cite{10.5555/3600270.3602070}, or how models perform in a “Zero-Shot” \cite{promptingguide2025fewshot} manner. 
\nocite{liao2023a3codgen,phan2024repohypersearchexpandrefinesemanticgraphs, wang2024rlcoderreinforcementlearningrepositorylevel, hai2025impactscontextsrepositorylevelcode, 10.1145/3691620.3695054, 10.1145/3643757}
CodeScholar \cite{shetty2023codescholar} addresses this by generating realistic, idiomatic API usage examples to help developers understand how to use unfamiliar libraries, serving as a complement to traditional documentation.  These examples not only improve human comprehension but also improve the quality of LLM-generated code when used in downstream tasks. We propose a specific prompting framework -– ReadMe.LLM —-which leverages assets (e.g. function signatures, examples, and descriptions) from a respective software library to assist code generation tasks. 

Additionally, there are different techniques for delivering updated contexts to an LLM. Recently, continual learning (CL) research has grown to be a good workaround to model cutoff dates. CL enables models to integrate new knowledge without forgetting past information through processes such as multiple training stages \cite{wu2024continual}. This shift underscores the importance of efficient mechanisms for integrating new knowledge. 

In parallel, tools have been developed to automatically extract information from GitHub libraries. \textit{Gitingest}, a popular tool, automatically extracts the repository directory structure and aggregates its files to be easily copied \cite{gitingest}. This enables users to easily copy entire repositories when trying to prompt LLMs. However, when applying this tool to our case studies, we found that the resulting file was too large and often caused the models to hallucinate. 

Libraries aren't the only tools that can be used as a building block by LLMs. A related approach to providing LLM-specific files is \textit{llms.txt}, a structured Markdown file organizing website content for LLMs and Agents \cite{llmstxt}, While web content is primarily designed for human users, this can be restrictive to LLMs with search capabilities or Agents that interact with the web. By providing a concise and structured representation of the content, \textit{llms.txt} can enhance usability. This takes inspiration from \textit{robots.txt} files, which detail which URLs a search engine crawler can access. In this context, our ReadMe.LLM proposal extends this idea by offering a well-defined framework for code generation tasks, as opposed to general website content.

\subsection{A Novel Elementary Approach: ReadMe.LLM}
Current documentation, such as ReadMe.md files, is written for human readers, but LLMs interpret information differently and are less effective with human-targeted formats. We argue that there should be LLM-targeted documentation \footnote{As we were writing this report after completing our experiments, Andrej Karpathy tweeted: "It's 2025 and most content is still written for humans instead of LLMs. 99.9\% of attention is about to be LLM attention, not human attention. E.g. 99\% of libraries still have docs that basically render to some pretty .html static pages assuming a human will click through them. In 2025 the docs should be a single your\_project.md text file that is intended to go into the context window of an LLM. Repeat for everything."\cite{karpathy2025tweet}.}. To address this, we propose ReadMe.LLM, a structured format to streamline library usage by LLMs: 
\begin{itemize}
    \item Optimized Documentation for LLMs: ReadMe.LLM provides structured descriptions of the codebase. Just as traditional header files help tell how to use a library to a traditional compiler, the ReadMe.LLM file tells an LLM how to effectively use this library to get things done. 
    \item Seamless Integration: Library developers easily create and attach a ReadMe.LLM to their codebase, which engineers can copy-paste or upload along with their query.
\end{itemize}
This approach shifts the focus to empowering libraries to be LLM-friendly, fostering adoption of emerging libraries. The overall workflow is illustrated below:

\begin{figure}[H]
    \centering
    \includegraphics[width=0.6\textwidth]{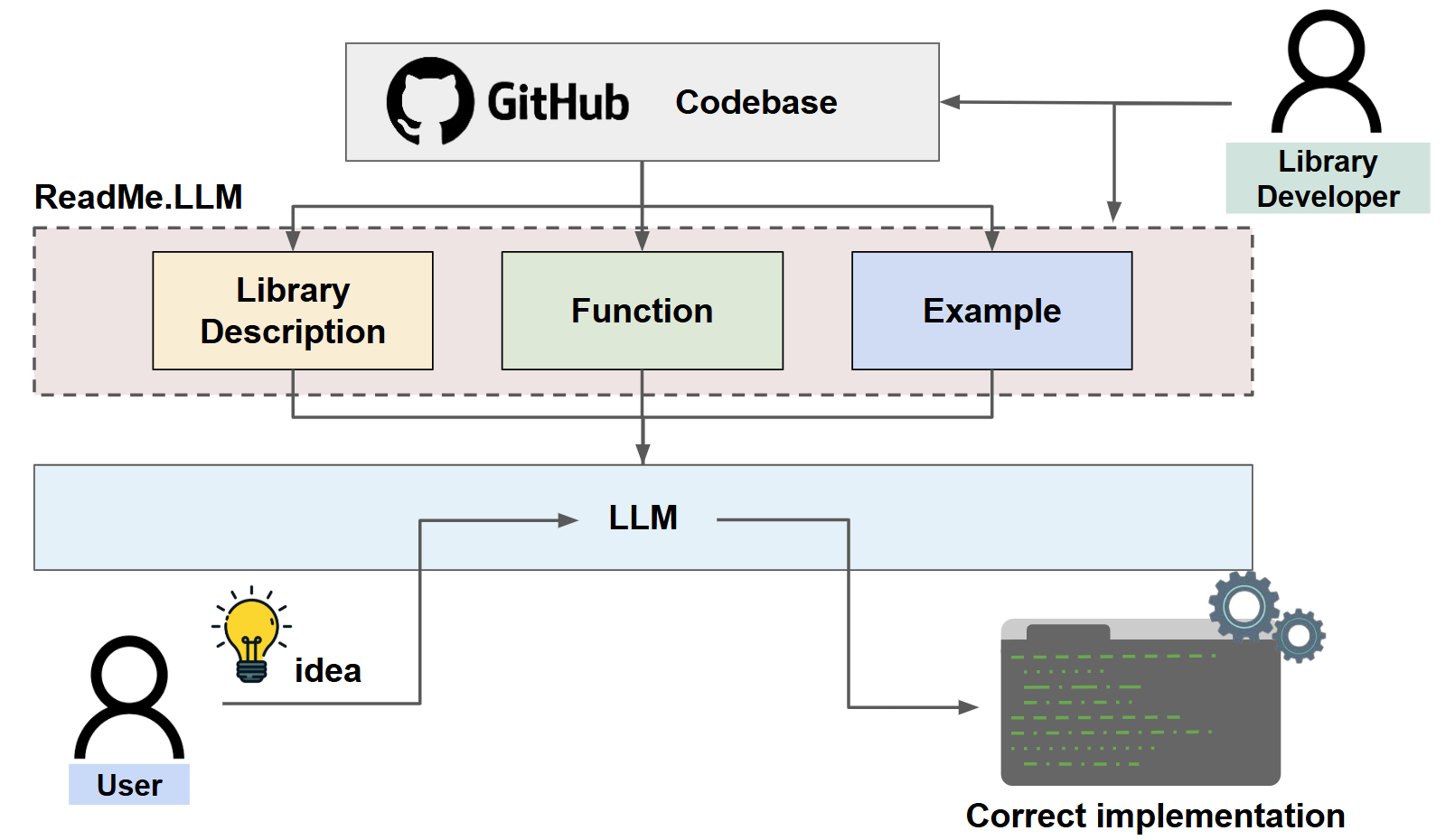}
    \caption{Depicting how ReadMe.LLM works}
\end{figure}

\section{ReadMe.LLM}
ReadMe.LLM is LLM-oriented documentation: a structured format leveraging assets from a software library to assist code generation tasks. Just as a ReadMe.md file provides essential information to human developers, a ReadMe.LLM provides it to LLMs. Based on our explorative testing and research into prompt engineering strategies, we propose the following ReadMe.LLM structure:
\begin{enumerate}
    \item Rules: A customizable set of guidelines that instruct the LLM on how to process the library’s information
    \item Library Description: A concise overview that sets the scene by outlining the library's purpose, core functionalities, and domain context.
    \item Code snippets: Clear function signatures are paired with illustrative examples that demonstrate real-world usage and expected outcomes.
\end{enumerate}
This was the structure we found worked best based on the libraries we experimented with; however, libraries from different domains may need to make small adjustments. We used XML tags to separate different types of content (e.g. $<$examples$>$). This formatting improves readability for LLMs and helps them easily parse the rules, description, and code snippets in ReadMe.LLM \cite{openai_text_api}. The complete ReadMe.LLM docs used in our experiments can be found in Appendix~\ref{app:readmellm_content}. In order to create a ReadMe.LLM for a library follow the instructions in Appendix~\ref{app:readmellm_instructions}.

\begin{figure}[H]
    \centering
    \includegraphics[width=0.45\textwidth]{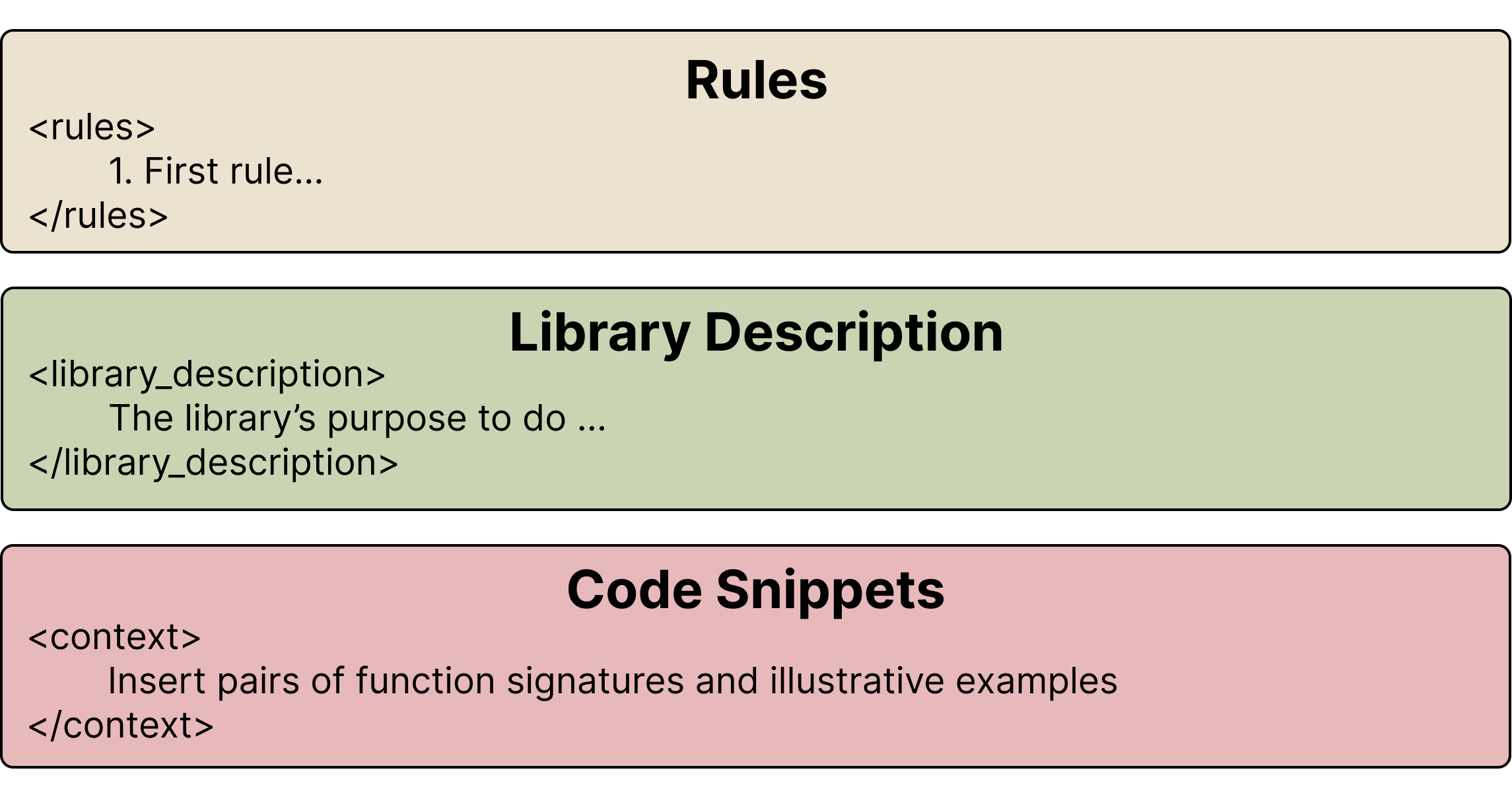}
    \caption{Example ReadMe.LLM Structure}
\end{figure}

With ReadMe.LLM defined, we outline how developers can utilize it in the software development process. Below are workflows we envision for three main user personas – library developers, engineers, and AI agents.

\subsection{Library Developer}

\begin{figure}[H]
    \centering
    \includegraphics[width=0.6\textwidth]{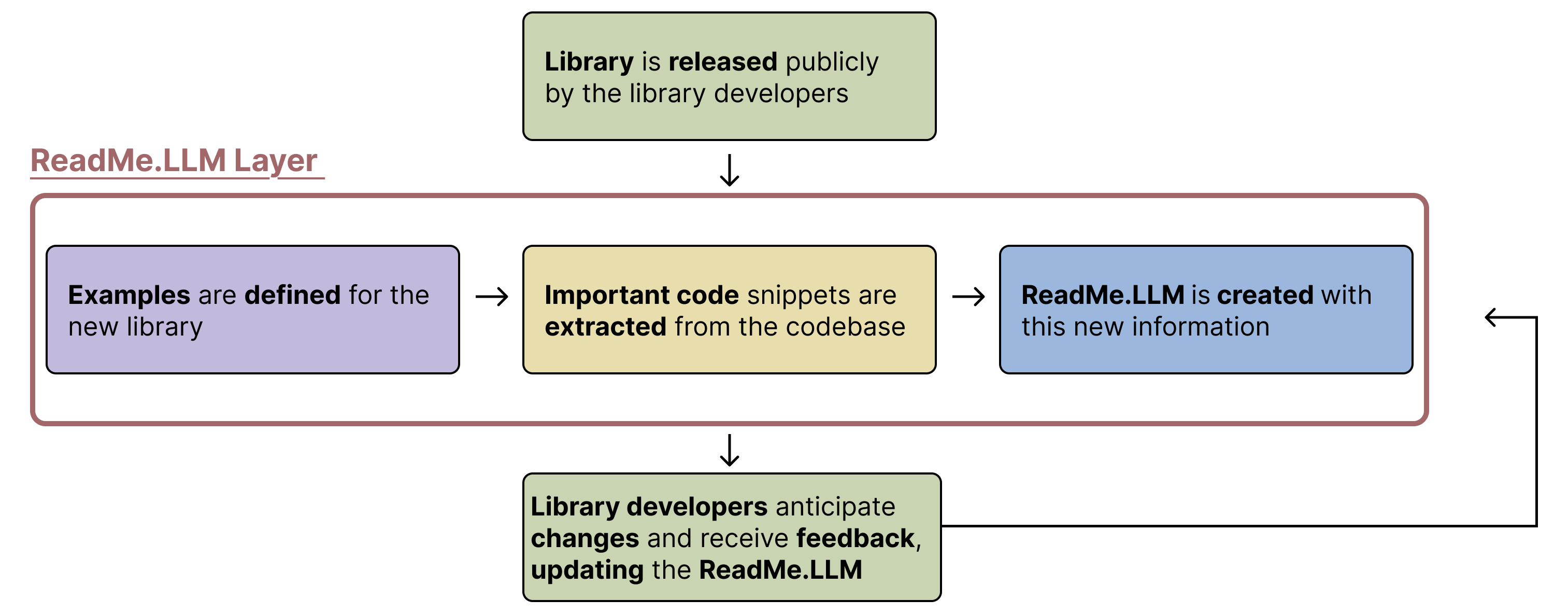}
    \caption{ReadMe.LLM integrated into library contributor workflow}
\end{figure}
Mirroring the ReadMe.md file for human documentation, we encourage library developers to create a ReadMe.LLM for their libraries to provide LLMs with targeted documentation that enhances coding outcomes. The general process is as follows: a library is released for users on Github, important code snippets and example usage are extracted from that codebase, and this is put together in a formatted text file-–the ReadMe.LLM. Once released, developers can engage with the user community to gather feedback and iterate on the ReadMe.LLM, improving its clarity and effectiveness over time. 
This falls into the software development cycle. Just as new releases for libraries require updated release notes to inform users of changes, library developers can edit the existing ReadMe.LLM file with any important changes.  

\subsection{Engineer}
\begin{figure}[H]
    \centering
    \includegraphics[width=0.6\textwidth]{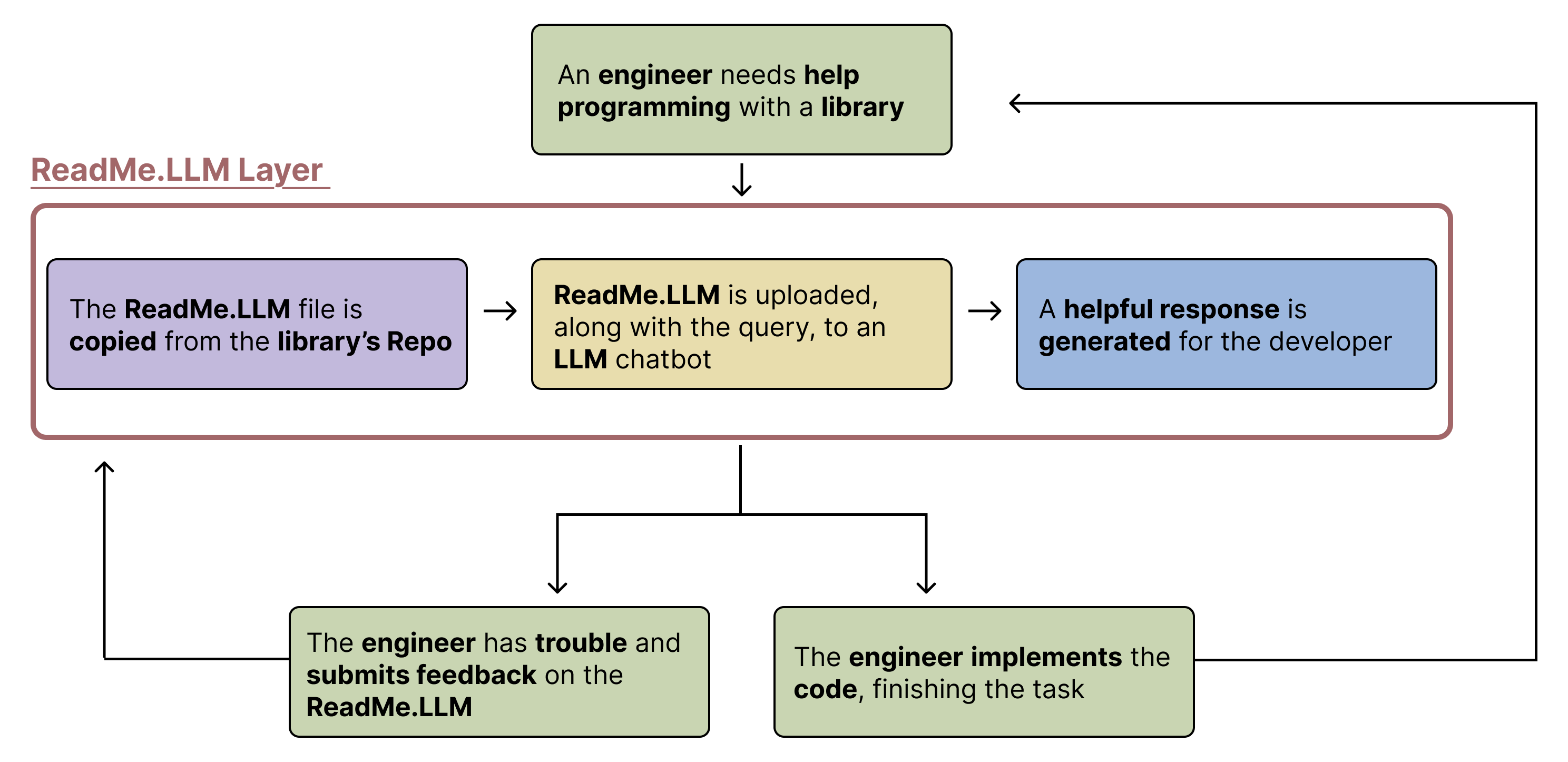}
    \caption{ReadMe.LLM integrated into engineer workflow}
\end{figure}
Many engineers have turned to an LLM for code generation assistance, but there is no standardized workflow for this process. We define the following general process: the engineers copy the ReadMe.LLM from the library’s repository, they then paste it into an LLM, and finally enter their query. With better context, the LLM provides more accurate and relevant code. If the engineer experiences any pain points, such as a missing function or an unclear example, they can submit this feedback to the library developers. 

\subsection{AI Agents}
AI Agents have become increasingly powerful and represent another user persona that can leverage ReadMe.LLM. Agents are powered by LLMs to automate tasks. Model Context Protocol (MCP) defines a standard way for AI agents to connect to data sources \cite{li2025deepdive, anthropic2024modelcontext}. MCP makes it easier for AI agents to process information and use different information sources when executing a task. 
Similarly, there has been a rise in the AI Agent libraries and services themselves, which use protocols such as MCP. Browser-Use \cite{browseruse2025repo} is a service that enables AI agents to automate tasks within web environments, and Manus \cite{manusim2025website} is an AI agent that executes tasks autonomously. With AI protocols such as MCP to manage the flow of tasks, Browser-Use and Manus can interact with each other and other tools more efficiently. 
Agents built using an MCP framework can seamlessly integrate ReadMe.LLM, allowing them to prioritize its contents, maintain context history across different ReadMe.LLM files, and navigate repositories efficiently. With this capability, agents can quickly locate and leverage the relevant ReadMe.LLM when tasked with coding. The process unfolds as follows: the AI Agent identifies a library to use for a task, locates the ReadMe.LLM file and processes it, combines this information with other sources to generate code, and finally, the agent debugs and optimizes the implementation before delivering the final output. 
This creates a more robust ecosystem where both human engineers and AI agents can utilize diverse libraries with ReadMe.LLM. 

\section{Experiments}
To understand what would be needed in a ReadMe.LLM, we systematically evaluated code that was generated by LLMs using different combinations of software library information. The software library information that we used includes human documentation (ReadMe.md files) and direct code snippets (full-function implementations, usage examples). 
To ensure robustness, we tested this across five different LLMs, all accessed through Perplexity: \textbf{GPT-4o}, \textbf{Sonar Huge (built on top of LLaMA 3.3 70B)}, \textbf{Claude 3.7 Sonnet}, \textbf{Grok-2}, and \textbf{Deepseek R1}. However, during our experimentation, DeepSeek R1 was temporarily removed from Perplexity, so we completed its testing via the DeepSeek website.

Something to consider is that LLMs have a knowledge cutoff, meaning they lack awareness of new information beyond their last training date (Table 1), and high training costs prevent frequent updates \cite{kumar2025llmpoweredclinicalcalculatorchatbot}. Since large-scale continual learning is still an open challenge, we relied on web search–which aggregates information from a broad range of sources \cite{perplexityai2025howwork}—as a practical alternative for accessing up-to-date information. This reflects realistic scenarios where users seek the most current insights. Additionally, to evaluate ReadMe.LLM’s utility in settings where web search is not feasible, such as internal company libraries, we included iterations without web search. 

\begin{table}[H]
\centering
\begin{tabular}{|l|l|}
\hline
\textbf{Model} & \textbf{Cutoff Date} \\
\hline
GPT-4o & October 2023 \cite{openai2024gpt4o, docsbot_models}\\
Llama 3.3 70B & December 2023 \cite{grattafiori2024llama3herdmodels, docsbot_models}\\
Claude 3.7 Sonnet & April 2024 \cite{anthropic2025claude37, docsbot_models}\\
Grok-2 & July 2024 \cite{xai2024grok2, xai2025modelspricing} \\
Deepseek R1 & July 2024 \cite{deepseekai2025deepseekr1incentivizingreasoningcapability, knostic2025deepseek}\\
\hline
\end{tabular}
\caption{Model Cutoff Dates}
\end{table}

With these LLMs, we experimented with two distinct libraries: DigitalRF \cite{DigitalRF2.6.11} and Supervision \cite{Roboflow_Supervision}. DigitalRF, an academic library with limited documentation, represents libraries that are not likely to have been included in the LLMs’ training process. Supervision, a modern, industry-run library, helps assess whether similar limitations persist for newer but more widely used libraries. 

For each library, we designed tasks based on consultations with the library developers to get insight into common use cases, ensuring LLMs interact with them realistically. We then provided these tasks to the LLMs, collected their generated code, and evaluated performance using two criteria: 
\begin{enumerate}
    \item Minimal Debugging – Code should work with at most three debugging rounds; the user pastes the error and the LLM regenerates fixed code based on that.
    \item Correct Library Utilization – The LLM should use the intended library functions rather than recreating functionality from scratch.
\end{enumerate}
After evaluating which context combinations yielded the highest success, we developed an optimal ReadMe.LLM for each library that generalizes well. We verified our optimal ReadMe.LLM on a held-out test using two previously untested LLMs –\textbf{Gemini 2.0 Flash} and \textbf{Mistral Large} (Cutoff dates: Dec 2024  \cite{docsbot_models} and 2023  \cite{neuroflash2025lechat}, respectively).

\subsection{Finding the Optimal ReadMe.LLM}
\subsubsection{Case Study 1: Supervision}
Supervision \cite{Roboflow_Supervision} is an industry-led library developed by Roboflow, which simplifies the process of working with computer vision models. It offers connectors to popular model libraries, a plethora of visualizers (annotators), powerful post processing features, and an easy learning curve. Main capabilities include: Detect and Annotate, Save Detections, Filter Detections, Detect Small Objects, Track Objects on Video, and Process Datasets. 

\textbf{Experiment Process}

For Supervision, we tasked LLMs with detecting, annotating, and cropping cars in an image. We selected an image with multiple objects (such as people, cars and buildings) to introduce complexity and tested the LLMs’ ability to generate code that differentiates between relevant and irrelevant detections. 

\begin{figure}[ht]
    \centering
    \includegraphics[width=0.8\textwidth]{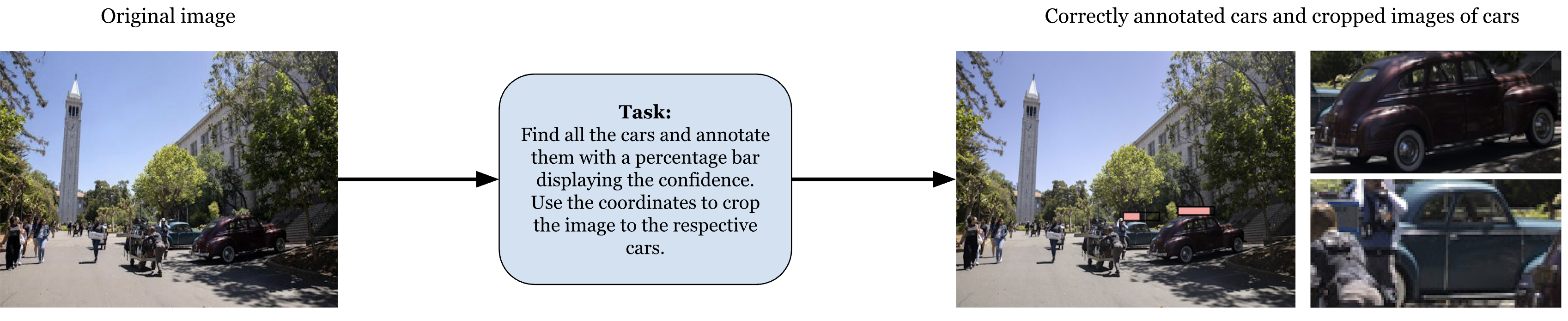}
    \caption{Supervision's Task 1 Case Study Summary}
\end{figure}

The LLM had to identify all cars, add a confidence score annotation, save the bounding box coordinates, and crop each detected car. To meet the Correct Library Utilization we mentioned above, the LLM should use Supervision’s Detections, Annotators, and Image Utility classes and functions, rather than alternative libraries and methods. The code should return at least one cropped image of a car, and the annotated picture should show confidence through either a bar or a percentage. 

\textbf{Results}

Figure 7 highlights that adding any context significantly improves LLM performance. The baseline success rate without context averaged around 30\%. Interestingly, DeepSeek R1 saw a decrease in performance when only given ReadMe.md as context – a potential sign that LLMs do not respond well to human-facing documentation. Relying solely on examples achieved a 96\% average success rate, while incorporating combined contexts enabled all models to hit 100\%, with the exception of Grok-2, which performed notably worse.

\begin{figure}[H]
    \centering
    \includegraphics[width=0.82\textwidth]{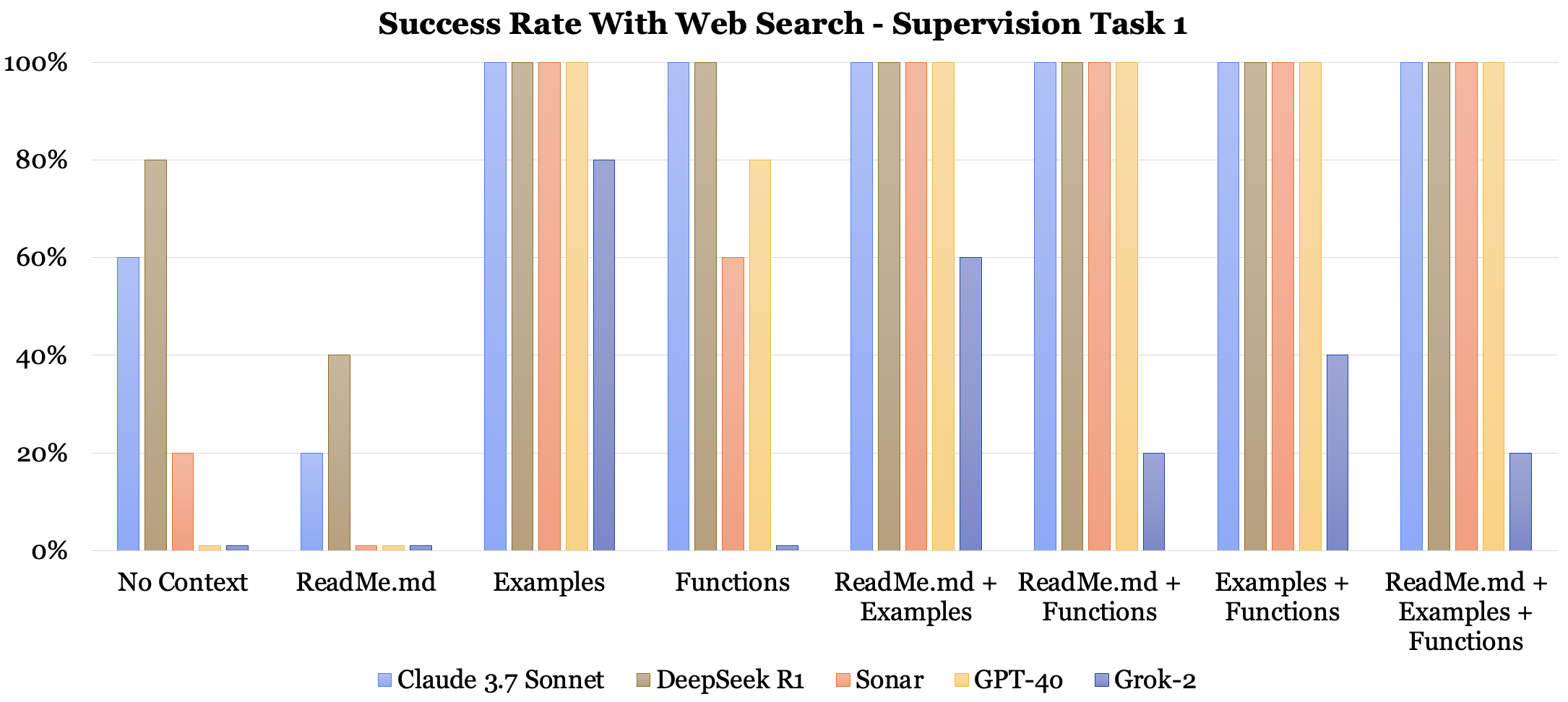}
    \caption{Supervision Task 1 Success Rates across various contexts and models}
\end{figure}


\subsubsection{Case Study 2: DigitalRF}
DigitalRF \cite{DigitalRF2.6.11} is an academic library developed by MIT Haystack that encompasses a standardized HDF5 format for reading and writing radio frequency (RF) data. Main capabilities include writing (converting an input WAV file into HDF5 format), and reading (converting HDF5 format back into a WAV file). 

DigitalRF presents an interesting contrast from Supervision. It is less popular and has minimal documentation. It is an example of a common class of libraries focused on file format translation, so testing it can help us assess the ReadMe.LLM idea in a broader context. 

\textbf{Experiment Process}

\begin{figure}[H]
    \centering
    \includegraphics[width=0.8\textwidth]{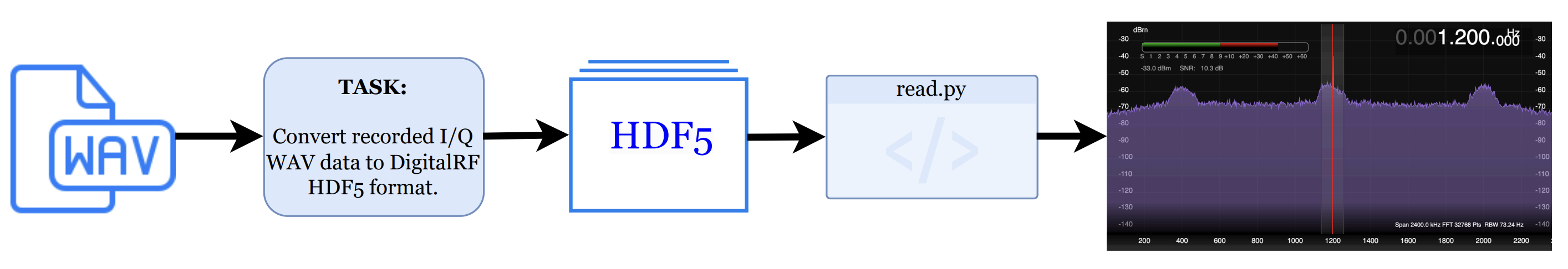}
    \caption{DigitalRF's Case Study Summary}
\end{figure}
For DigitalRF, we tasked the LLMs with writing a WAV file into DigitalRF-HDF5 format. We obtained a WAV file (a 10-second-long radio signal) containing I/Q data using a Software Defined Radio (SDR) and the SDR++ application, and tasked LLMs with converting it to a standardized HDF5 format using the DigitalRF library. To meet the correct library utilization requirement above, we made sure the LLM-generated code created a proper HDF5 folder structure. We ran this output through a pre-built script to reconstruct the original WAV file and ensured that it played back the original audio sample. 

\textbf{Results}

Similar to Supervision, we again see in Figure 9 that adding any context significantly improves LLM performance when generating code for unfamiliar libraries. The poor performance with ReadMe.md further proves that LLMs do not respond well to documentation that is intentionally made to be readable to humans. 

Incorporating structured information consistently led to better results. Among individual contexts, function-related information and examples had the strongest impact, both raising the average success rate to 64\%. For combined contexts, ReadMe.md + Functions achieved the highest success rate at an average of 88\%. 

\begin{figure}[H]
    \centering
    \includegraphics[width=0.82\textwidth]{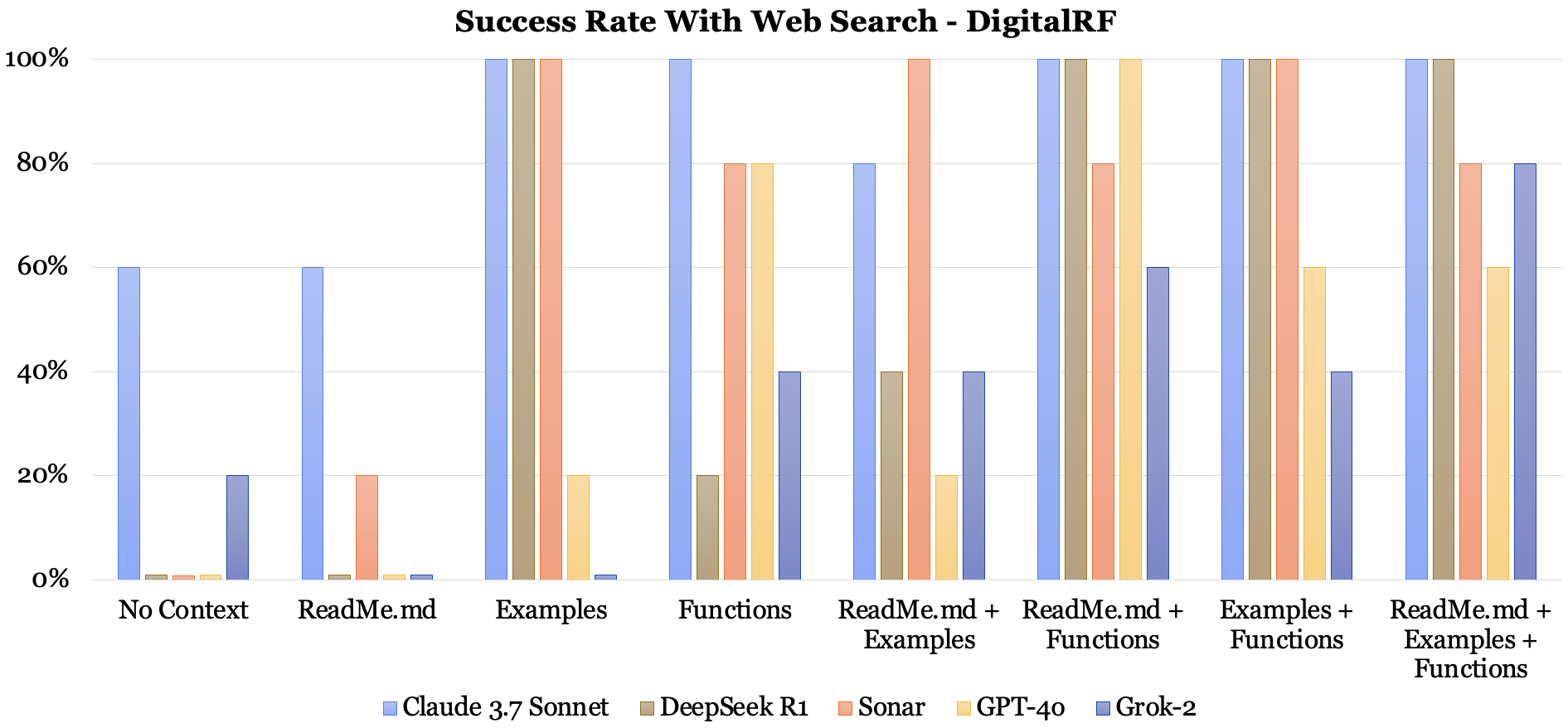}
    \caption{DigitalRF Task Success Rates across various contexts and models}
\end{figure}

\subsection{Verifying the  ReadMe.LLM}
After analyzing the case study results presented earlier, we initiated the design of an optimal ReadMe.LLM for both libraries.
The Supervision case study results revealed that using solely the ReadMe.md context led to lower accuracy than when no context was provided. Consequently, we decided to omit ReadMe.md information from our final ReadMe.LLM and instead included only code snippets—interweaving function implementations and code examples.

In the first version of Supervision’s ReadMe.LLM, we incorporated the complete Detections class, all Annotator classes, Image Utility functions, and corresponding examples. We tested this version against Sonar and Grok-2, the models that had previously underperformed. After several iterations, it became evident that this initial version’s extensive length led to hallucinations.

To reduce the length of the context, we revised the ReadMe.LLM by including examples and only function signatures, rather than full implementations. This final version achieved a 100\% success rate with Sonar and Grok-2; subsequent testing with GPT-4o, Claude 3.7 Sonnet, and DeepSeek R1 also yielded perfect performance when web search was enabled. When evaluated without web search, the ReadMe.LLM maintained this performance across all models, with the exception of Grok-2, which achieved an 80\% success rate. 

\begin{figure}[H]
    \centering
    \includegraphics[width=0.55\textwidth]{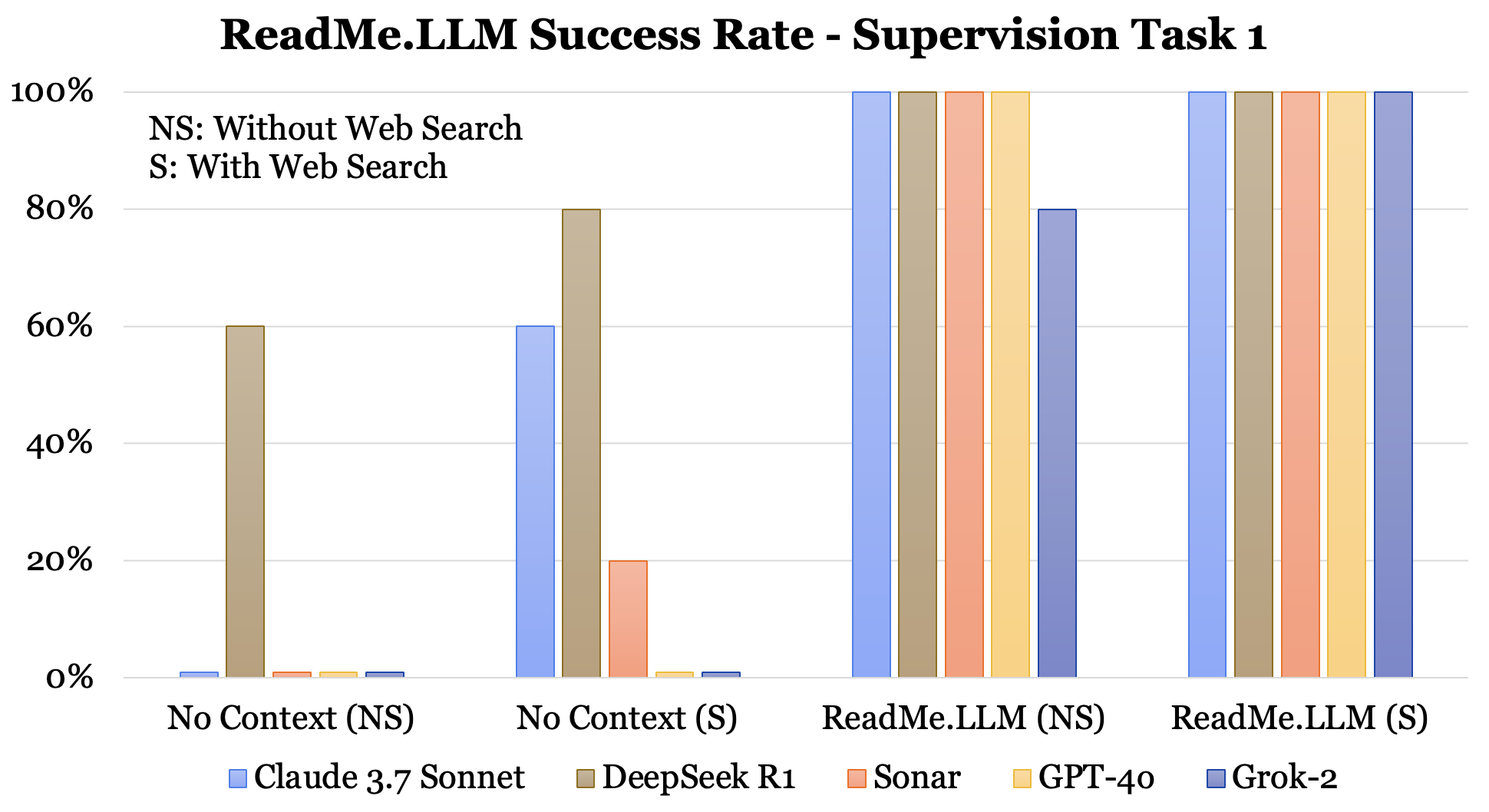}
    \caption{Supervision's ReadMe.LLM Success Rates for Task 1 across various models}
\end{figure}

To verify ReadMe.LLM generalizes to other Supervision use cases, we designed a second task. In this task, the LLM was required to identify individuals within an image, apply a blur to each person, and overlay a different image on each subject.  
\begin{figure}[H]
    \centering
    \includegraphics[width=0.8\textwidth]{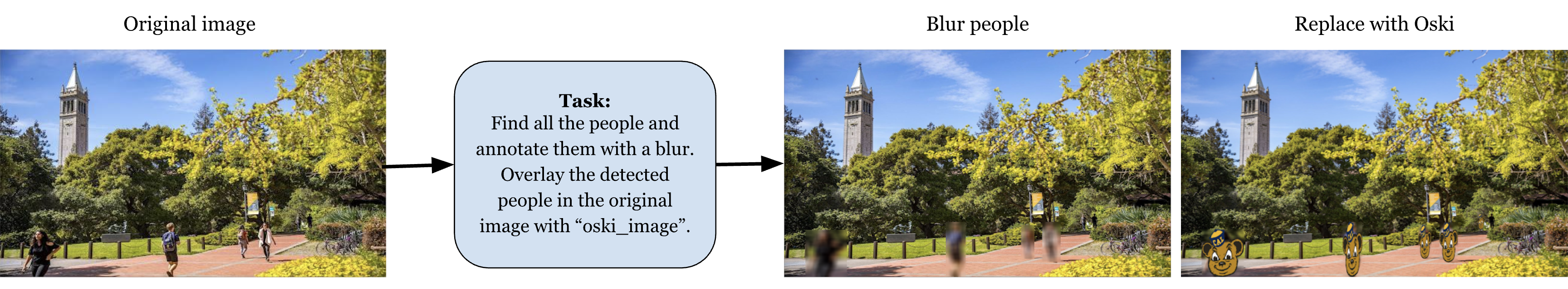}
    \caption{Supervision's Task 2 Case Study Summary}
\end{figure}
As shown in Figure 12, all LLMs performed poorly in zero-shot coding—even with web search activated—with only DeepSeek R1 occasionally succeeding. However, when the ReadMe.LLM was provided, the success rate increased to 100\% across all models, demonstrating its adaptability to a variety of tasks.  

\begin{figure}[H]
    \centering
    \includegraphics[width=0.55\textwidth]{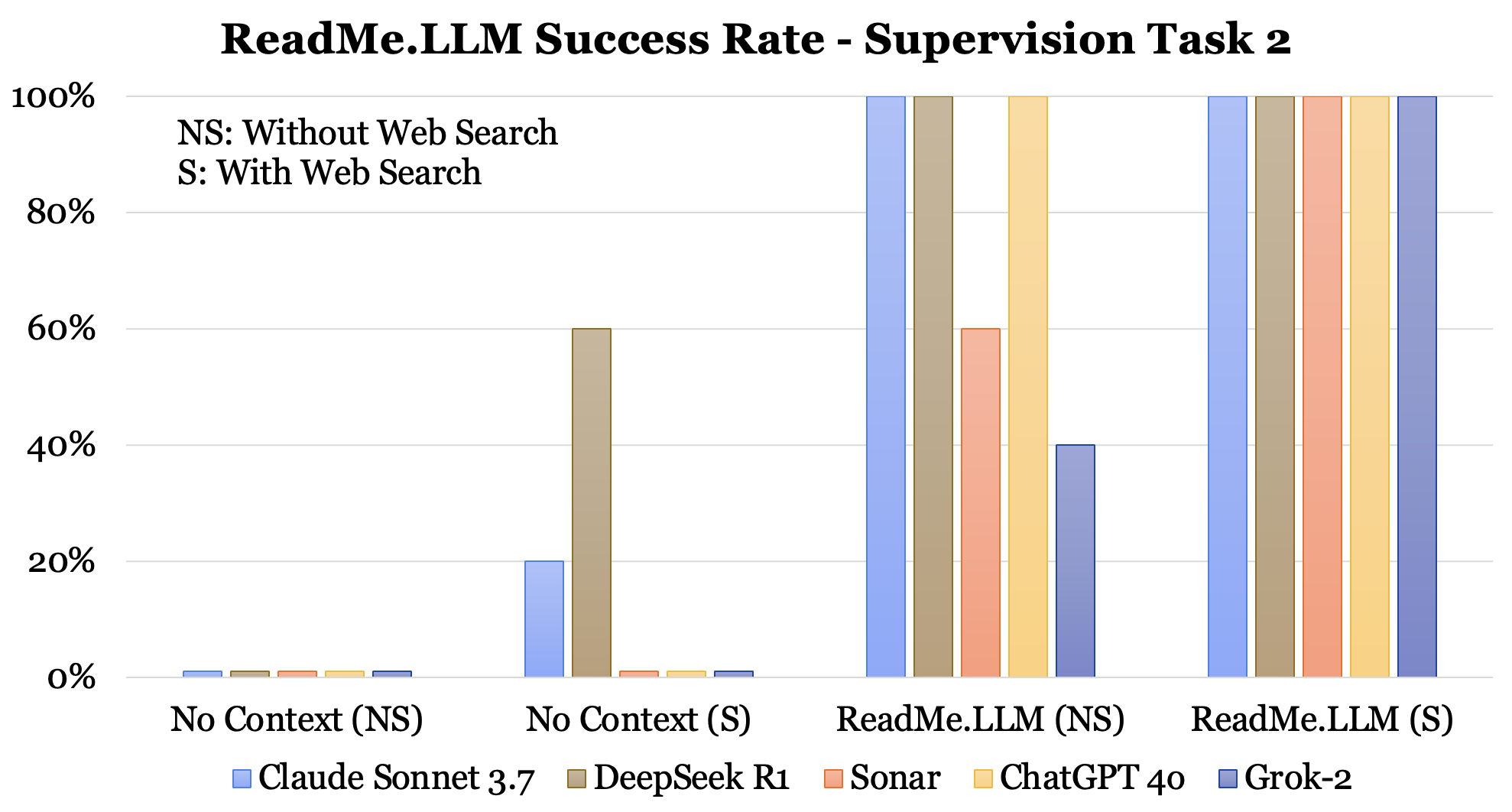}
    \caption{Supervision's ReadMe.LLM Success Rates for Task 2 across various models}
\end{figure}

Subsequently, we employed an analogous approach to construct DigitalRF’s ReadMe.LLM, directly interweaving function signatures and examples from the repository, similar to our process for Supervision. This approach immediately yielded a 100\% success rate for Sonar and Grok-2 when web search was enabled, and was therefore adopted as our final ReadMe.LLM for DigitalRF. Among the remaining three models, only GPT-4o did not achieve perfect performance with web search, attaining only 80\%. When web search was disabled, the average success rate dropped to 70\%, with only DeepSeek R1 maintaining a 100\% success rate. These results suggest that further refinements could yield an even more effective ReadMe.LLM for DigitalRF in future iterations.

\begin{figure}[H]
    \centering
    \includegraphics[width=0.57\textwidth]{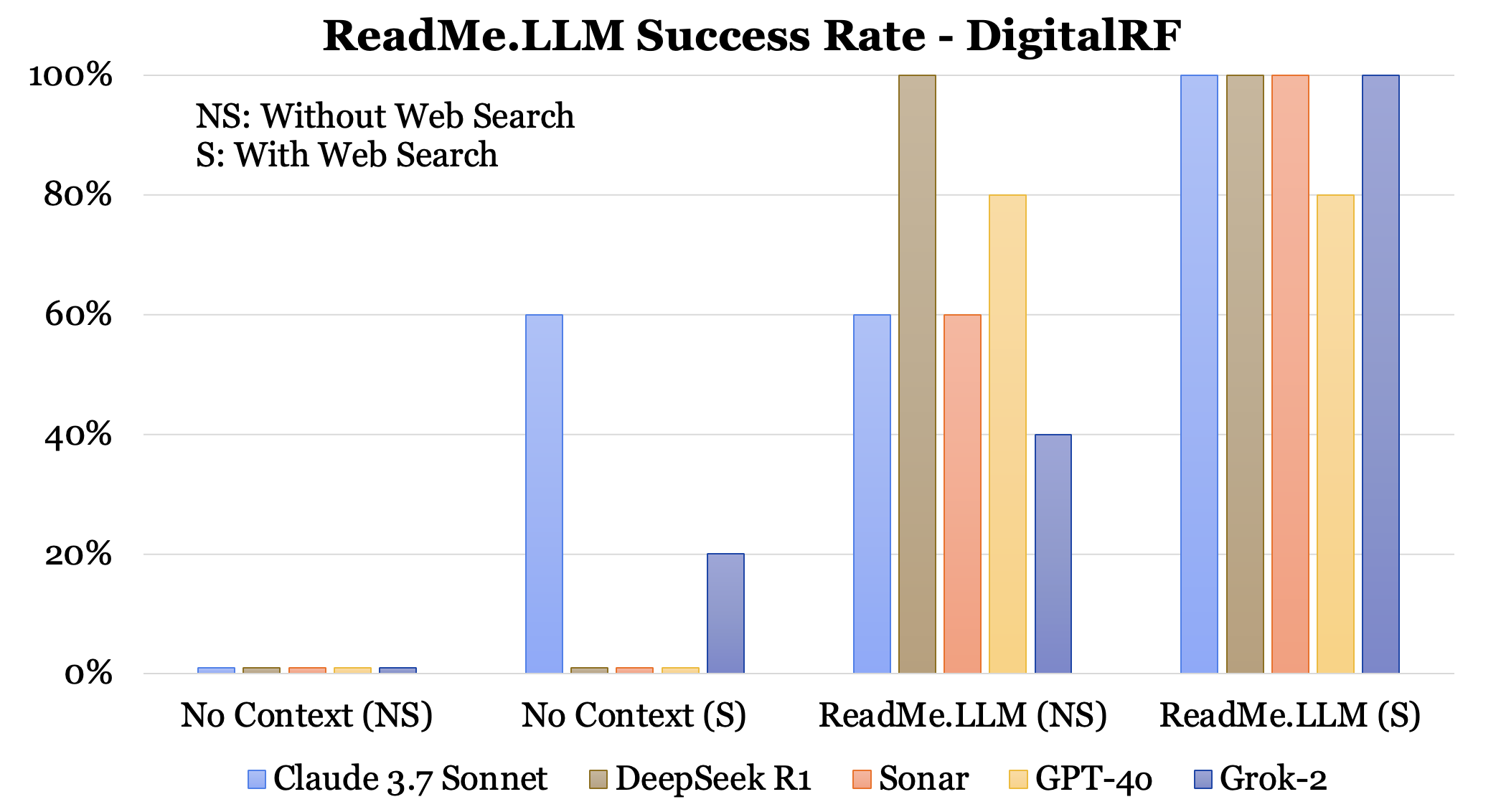}
    \caption{DigitalRF's ReadMe.LLM Success Rates across various models}
\end{figure}

A comparison of results for earlier model versions across all contexts is provided in Appendix~\ref{app:appendix_older_models}

\subsection{Held Out Tests}
To further assess the robustness of ReadMe.LLM, we conducted a held-out test using three new models that had not been used in previous experiments: Gemini 2.0 Flash, Mistral Large, and GPT-4.1. This evaluation aimed to verify the effectiveness of our final framework when applied to LLMs that did not contribute to our development process. 

We began with Supervision. Even with web search enabled, zero-context prompting performed poorly. With ReadMe.LLM Gemini’s success rate jumped to 100\% on the first task and 80\% on the second. Mistral, which had previously failed both, reached a perfect 100\% on both tasks, as did GPT-4.1. Without web search, Supervision's ReadMe.LLM sustained a high success rate with these models, with only Gemini exhibiting a slight decline.

\begin{figure}[H]
    \centering

    \begin{subcaptionbox}{Supervision Task 1\label{fig:sub1}}[0.45\textwidth]
        {\includegraphics[width=\linewidth]{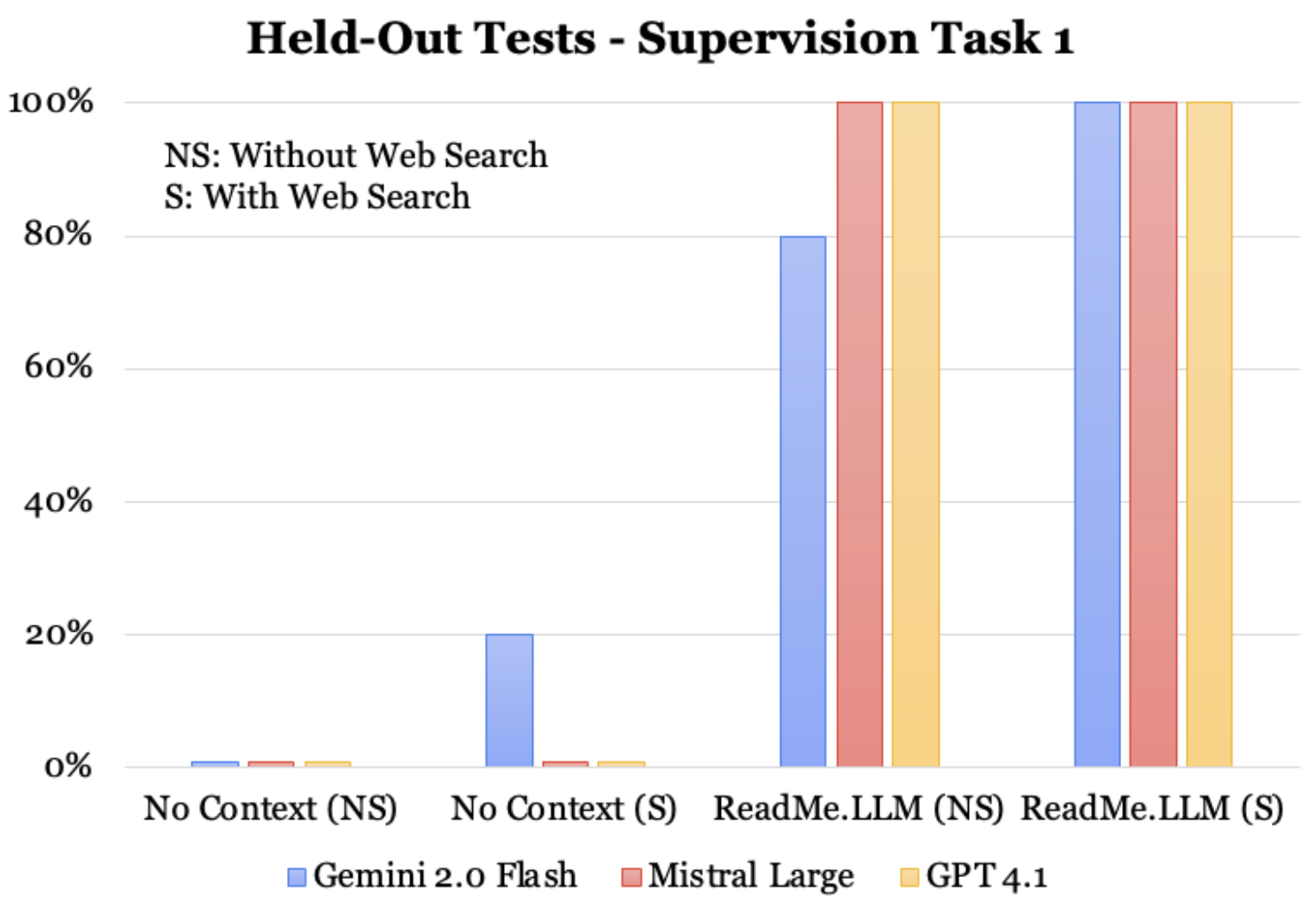}}
    \end{subcaptionbox}
    \hspace{0.05\textwidth}
    \begin{subcaptionbox}{Supervision Task 2\label{fig:sub2}}[0.45\textwidth]
        {\includegraphics[width=\linewidth]{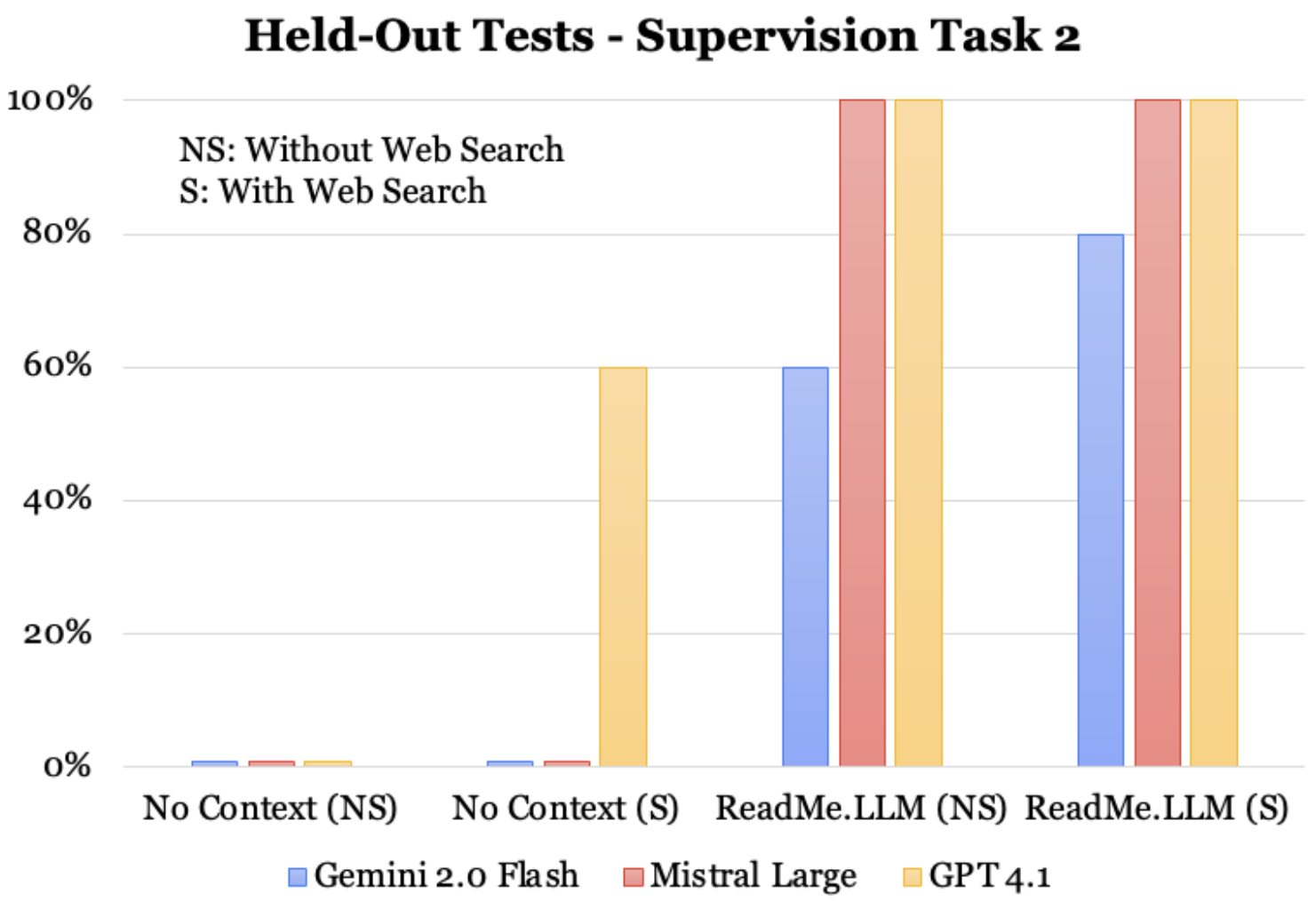}}
    \end{subcaptionbox}

    \caption{Supervision's Held-Out Tests}
    \label{fig:held_out_supervision}
\end{figure}

We tested these same models with DigitalRF. When prompted without any additional context, accuracy was 0\%, even with web search capabilities. However, once ReadMe.LLM was applied, Gemini and Mistral achieved a 80\% success rate and GPT-4.1 100\%, showing a dramatic and consistent improvement. In line with the original models, the absence of web search capabilities resulted in a slight performance decline with Gemini and Mistral, but still remained significantly superior to conditions without context.

\begin{figure}[H]
    \centering
    \includegraphics[width=0.48\textwidth]{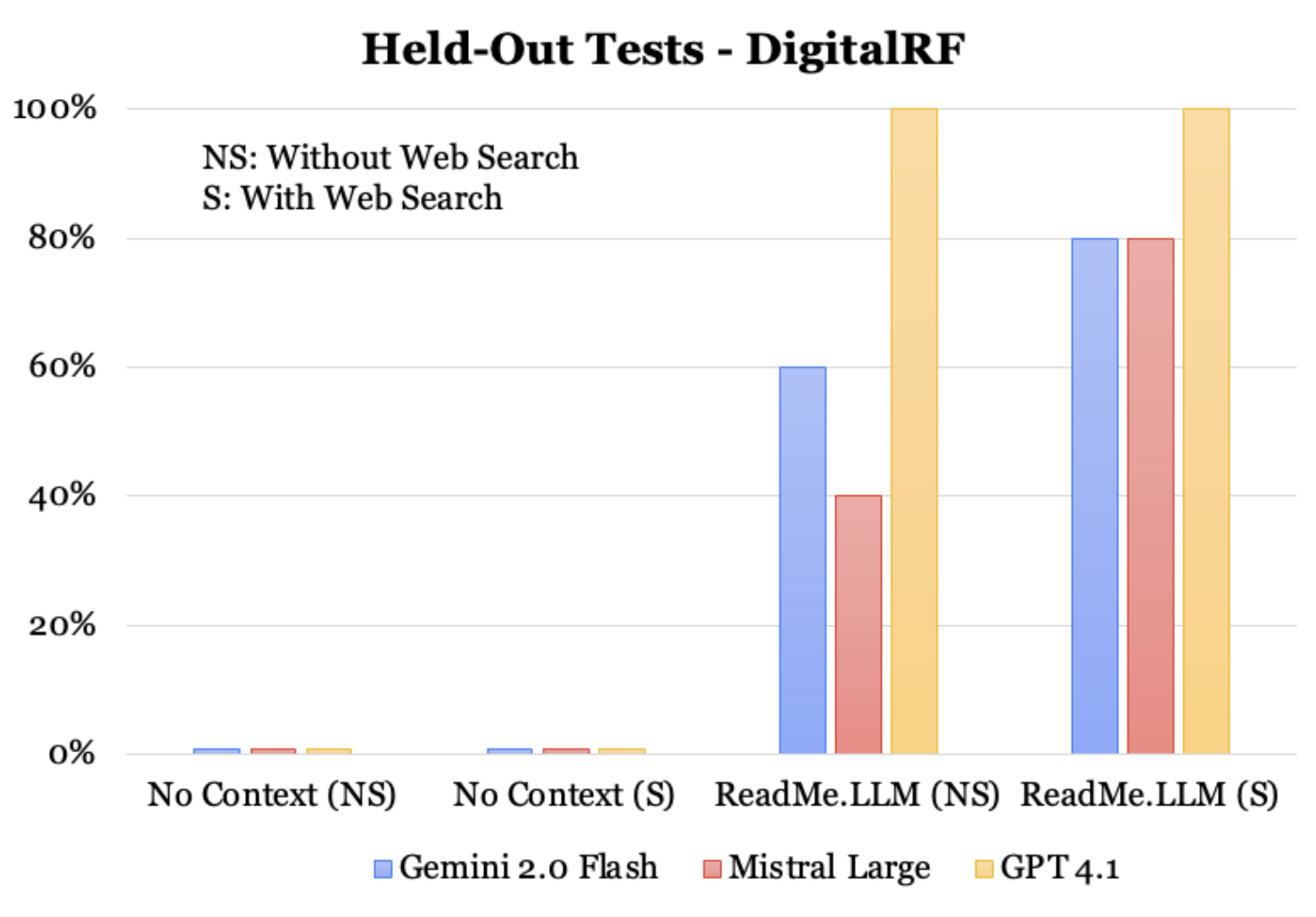}
    \caption{DigitalRF's Held-Out Tests}
\end{figure}

This consistent performance boost demonstrates that ReadMe.LLM significantly enhances code generation capabilities, even for held-out models and libraries that did not influence the development of ReadMe.LLM. Our results affirm that ReadMe.LLM not only improves accuracy with familiar models, but also generalizes effectively to entirely new ones. By bridging gaps in LLM knowledge, ReadMe.LLM makes code generation more reliable and robust across diverse architectures and domains. 

\section{Discussion}
\textbf{It’s Possible to Seamlessly Improve Code Generation Through Prompting with LLM-Oriented Documentation.} Through our experimentation and research, we show that prompting with LLM-oriented software library documentation —ReadMe.LLM– can greatly increase the performance of code completion tasks for LLMs. ReadMe.LLM can perform optimally in the majority of scenarios irrespective of model selection. This is a powerful tool for library developers and engineers, because it increases the accessibility and performance of leveraging LLMs for code completion. Engineers no longer have to create complex prompts, use computationally-intensive methods like RAG, or drastically change their queries. Instead, they can just copy and paste a library’s ReadMe.LLM into an LLM's prompt window and give a code generation task as usual \footnote{After our paper was first released, Andrej Karpathy tweeted in support of the seamless experience of copy and pasting documentation for LLMs: "Tired: elaborate docs pages for your product/service/library with fancy color palettes, branding, animations, transitions, dark mode, … Wired: one single docs .md file and a 'copy to clipboard' button" \cite{karpathy2025vibecoding}.}.
Similarly, library developers can develop a ReadMe.LLM to ensure that their library is being correctly represented by an LLM and therefore, seamlessly used by the targeted engineer. 
 
\textbf{Tailoring context selection to the model can improve code quality.} Through our experimentation, we have found that different models have varying success with diverse contexts. While we have identified a framework that consistently performs well across tested models, there may be situations where you can gain even better performance by modifying the library’s ReadMe.LLM. For example, in our DigitalRF case study, we observed that Sonar achieved a 100\% success rate with ReadMe.md and Examples, but dropped to 80\% with ReadMe.md and Functions.  However, the other models (Grok-2, GPT-4o, Claude 3.7, and DeepSeek R1) saw better performance with ReadMe.md + Functions. With this in mind, it may be advisable for library developers to test their ReadMe.LLM against a wide variety of LLMs to ensure its robustness.

\textbf{Patterns in Models’ Limitations.} There are several challenges an LLM would face when completing our tasks. After our experimentation, we were able to categorize these challenges into LLM code generation insights. 

First, it became a common occurrence that a model would fail on a task, not because of the usage of the target library, but because of the prerequisite Input/Output tasks. Errors such as importing a suitable library, creating a new file to save data, or reading the correct data often caused the task to fail. For example, with DigitalRF, the model failed at reading in the WAV file because the LLM-generated code utilized the wrong Python library that supported a different file format. We argue that this is a general reflection of an LLM’s ability to generate code for IO-related tasks. With this, if a library wants to enhance developer use, developers should provide necessary context about IO-tasks in the ReadMe.LLM. 

Second, some models are able to complete the task but do not use the functions provided by the specified library. We marked these occurrences as failures, since utilizing the actual library functions is typically much more efficient than assembling a solution from a mix of other libraries. For example, during our Supervision tasks, the library offered functions to crop or overlay an image, but the LLM generated code that performed these tasks using the CV library instead, resulting in much longer and less efficient code. This is also important from a Software Bill of Materials (SBOM) \cite{10.1109/ICSE48619.2023.00219} perspective: an SBOM is an inventory of all software components and dependencies, and using the correct library functions ensures that the SBOM accurately reflects the software’s dependencies, improving transparency and security. This highlights the importance of providing clear examples and function definitions, especially when the desired functionality may overlap with existing libraries that a model is trained on. Doing so can help guide the model to use the correct function from the targeted library.

\textbf{LLM Performance is a Moving Target.}
While our final results demonstrate consistent gains using ReadMe.LLM, earlier versions of the same models --  shown in Appendix A -- showed far more dramatic improvements.

This raises a broader question: as models evolve, which challenges are temporary and which are more fundamental? For example, hallucinations might fade as models improve, or they might persist in new forms. Recent work from Nexusflow on Athene-V2 suggests that models often face a tradeoff: being fine-tuned for conversational chat vs. agentic behavior \cite{nexusflow2024athenev2}. This specialization means models may not be able to excel at all tasks simultaneously. And regardless of progress, smaller local models will likely continue to face more limitations due to resource constraints. All of this reinforces the idea that LLM evaluation must keep pace with a fast-moving target.

\textbf{Jailbreaking and Prompt Robustness.}
Recent safety research underscores that LLM behavior can be dramatically altered through structured prompting alone. For instance, HiddenLayer’s KROP technique bypasses safety safeguards simply by reformatting the input using specific structural patterns, without needing to inject unsafe content \cite{hiddenlayer2025krop}. Inspired by this behavior, ReadMe.LLM encloses library information in XML tags -- a way that LLMs are highly responsive to. However, this strength may also be a weakness: such formatting has recently been characterized as a form of "jailbreaking," since many models appear overly obedient to syntactically clean inputs like XML. If future alignment interventions are trained to block these structured jailbreaks, ReadMe.LLM’s effectiveness could degrade -- even if its content remains entirely safe and constructive. This highlights the need to monitor prompt performance over time and investigate how safety tuning may unintentionally interfere with developer-facing tools.

\textbf{Similar Approaches in Practical Evaluation.} The Berkeley Function-Calling Leaderboard has implemented data filtering and quality enhancement techniques when using live, user-contributed function documentation and queries to benchmark LLM performance in function calling \cite{berkeley-function-calling-leaderboard, patil2023gorillalargelanguagemodel}. Their definition of high-quality documentation—structured JSON containing key fields such as function name, description, and detailed specifications—closely aligns with the design of ReadMe.LLM, proving the generalizability of our structure. Furthermore, their evaluation includes comprehensive benchmarking across programming languages and libraries, incorporating methods such as function relevance detection and Abstract Syntax Tree (AST) analysis. Integrating these evaluation techniques into the ReadMe.LLM assessment could enhance its robustness to benchmark and highlight its practical benefits.

In effect, the new thing we are advocating is that every library developer should learn from the benchmark design, just as LLM developers learn from benchmarks. Knowing that LLM developers are chasing higher benchmark scores also informs  library developers. While perhaps large popular libraries have the luxury of expecting LLM-providers to optimize for them, smaller niche libraries would do well to make themselves look like the benchmarks that LLM providers optimize for.

\textbf{Extending ReadMe.LLM to Tool Use.} Recent work has increasingly focused on LLMs' ability to interact with external tools, such as invoking APIs or executing functions. As LLM applications become more powerful and autonomous, their ability to correctly use APIs becomes critical. Similar to software libraries, APIs define the correct way to access desired functionality through structured calls. Inspired by this similarity, we propose extending ReadMe.LLM to support tool use by generating LLM-targeted documentation for APIs. Specifying important details like endpoints, parameters, or expected responses can be formatted and optimized for LLMs. Notably, the API-Bank benchmark demonstrates the importance of LLMs understanding API structure and semantics to improve tool use accuracy \cite{li2023apibankcomprehensivebenchmarktoolaugmented}. ReadMe.LLM can bridge the current performance gaps by providing LLM-friendly API documentation.

\section{Conclusion and Future Work}
We present ReadMe.LLM, novel LLM-oriented documentation that provides relevant context about a software library to assist code generation. We evaluated different combinations and structures of context and tested these across the current leading LLMs. We presented the optimal ReadMe.LLM structure, which has the highest average accuracy across different models, and increases correctness up to 100\%. 

As engineers continue to turn to LLMs when facing roadblocks, library developers must make their content easily available and understandable to LLMs. Failure to do so will not only hinder engineers by producing unreliable code but also disadvantage smaller libraries, perpetuating a cycle of underutilization and inefficiency. ReadMe.LLM becomes essential for bridging the gap between library documentation and Generative AI assistance.

Looking ahead, we remain committed to enhancing this framework by exploring new components and optimizing the structure of context delivery. We are planning on investigating how this framework extends to tasks other than code generation, such as question \& answering and code debugging. Additionally, in this paper, we focus on LLM chatbots, but ReadMe.LLM can be extended to co-pilots as well. With the rise of vibe-coding and the adoption of products like Cursor \cite{cursor2023}, improving the code generation capabilities directly in the editor is important. A co-pilot could recognize the ReadMe.LLM file within an imported module’s directory and utilize it to generate more relevant and accurate code for its user. 

We welcome contributions from the community to advance this initiative and shape the future of LLM-library interactions. Please explore our website, \href{https://readmellm.github.io/}{readmellm.github.io}, and we encourage you to contribute to online discussions. 

\section*{Acknowledgments}
We thank the National Science Foundation and especially the SpectrumX (AST-2132700) community for its support. We also thank the UC Berkeley College of Engineering's Fung Institute, as well as Dr. Josh Sanz for helpful conversations.

\bibliographystyle{ieeetr} 
\bibliography{egbib} 

\clearpage
\appendix

\section{Appendix: Comparison of results with older models}\label{app:appendix_older_models}

\subsection{DigitalRF's Case Study with Older Models}\label{app:appendix_older_digitalRF}

In our original experiments on the DigitalRF case study, we evaluated Claude 3.5 Sonnet across all eight context combinations and ReadMe.LLM, with web search enabled. When Claude 3.7 Sonnet was released, we repeated the same experiments. 

A clear pattern emerges: while both versions benefit from richer context, the older Claude 3.5 model exhibited a larger performance jump when provided with our ReadMe.LLM versus human-oriented documentation. Specifically, Claude 3.5’s accuracy moved from 20\% (no context or ReadMe.md) to 80\% with ReadMe.LLM—a 60\% point gain—whereas Claude 3.7 improved from 60\% (no context or ReadMe.md) to 100\% with ReadMe.LLM—a 40\% point gain. This indicates that ReadMe.LLM had a greater impact in bridging the knowledge gap in the earlier version of the model. As models become more capable (e.g., Claude 3.7), the baseline performance on human-oriented docs rises, but ReadMe.LLM continues to provide a consistent boost.

\begin{figure}[H]
    \centering
    \includegraphics[width=0.6\textwidth]{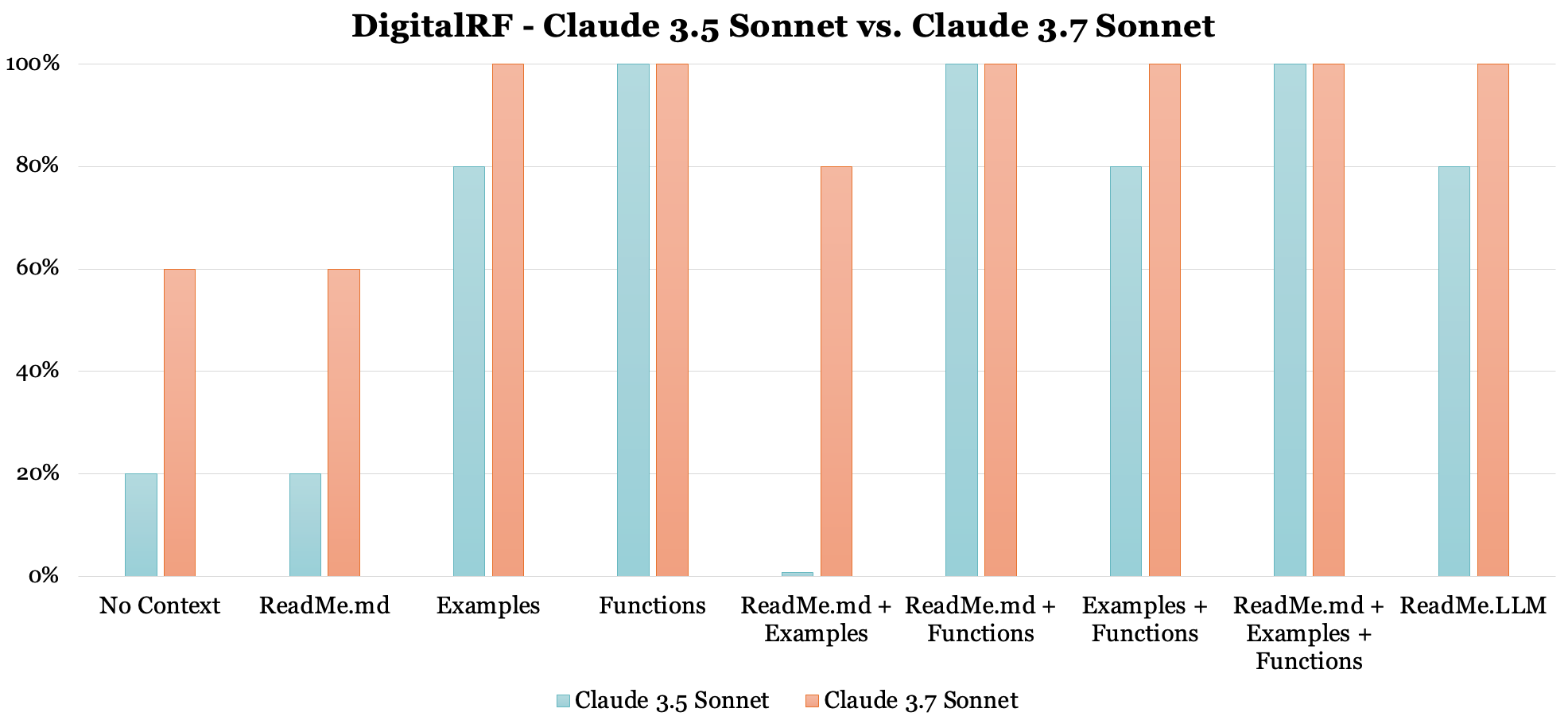}
    \caption{DigitalRF - Claude 3.5 Sonnet vs. Claude 3.7 Sonnet}
\end{figure}

When a new Sonar model was released built on Llama 3.3 70B, we also repeated our DigitalRF experiments to compare it with the older Sonar based on Llama 3.1 70B. Although we did not evaluate ReadMe.LLM on the older Sonar (Llama 3.1 70B), the eight standard context combinations reveal some clear trends. Both Sonar 3.1 and Sonar 3.3 achieved 100\% success when provided only with examples, underscoring the power of real-world snippets. Most notably, Sonar 3.1 hallucinated dramatically under richer prompts—success falls to 40\% with ReadMe.md + Examples and 20\% with the full mixed context—whereas Sonar 3.3 remained far more robust (100\% and 80\%, respectively). In both versions, ReadMe.md alone yielded only a 20\% success rate, again showing that human-oriented documentation offers minimal benefit.

\begin{figure}[H]
    \centering
    \includegraphics[width=0.5\textwidth]{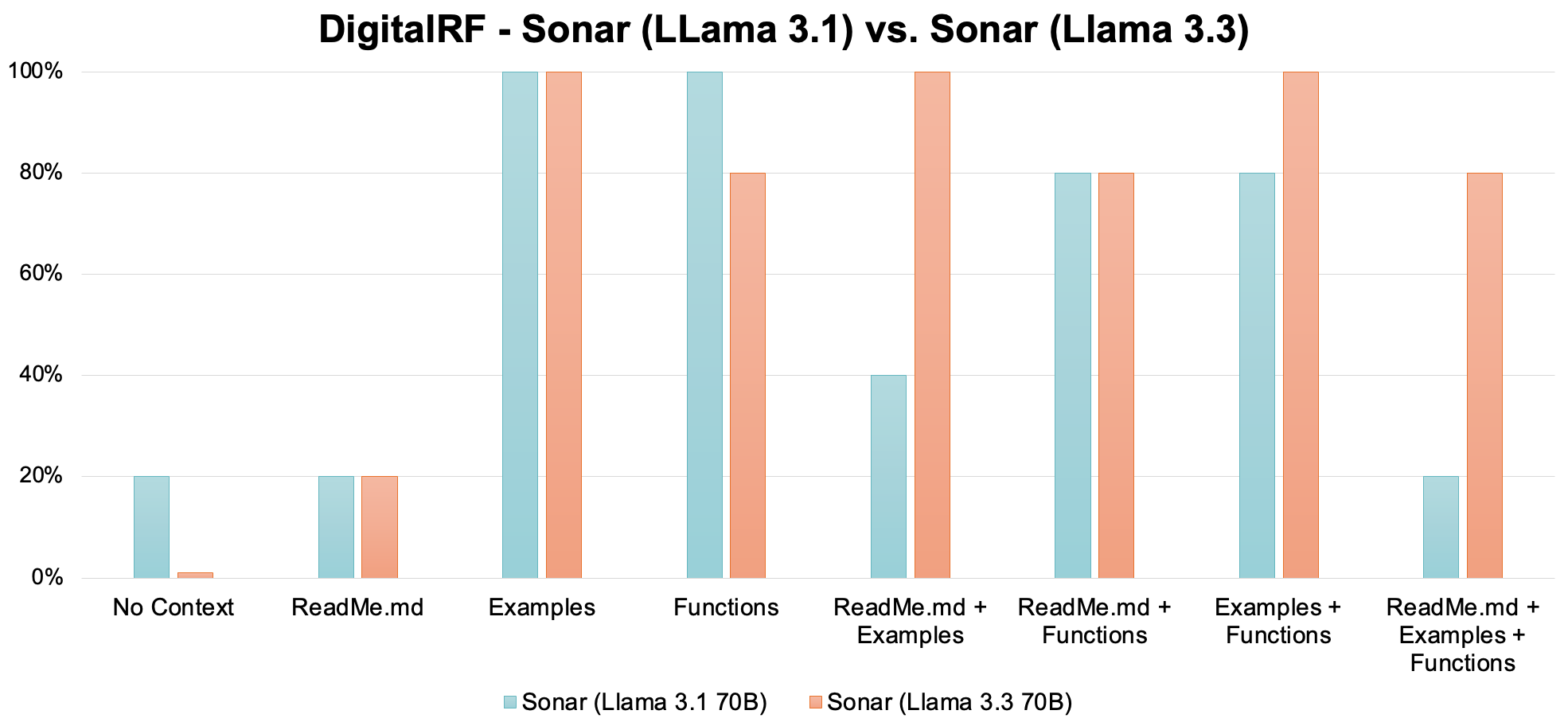}
    \caption{DigitalRF - Sonar (LLama 3.1) vs. Sonar (Llama 3.3)}
\end{figure}

\subsection{Supervision's Case Study with Older Models}\label{app:appendix_older_digitalRF}

When Claude was upgraded from 3.5 Sonnet to 3.7 Sonnet, we also reran Supervision's Task 1 case study experiments. Claude 3.5 failed always with no context or ReadMe.md alone, yet jumped to a perfect 100\% as soon as it saw examples, functions, or any combined context—including ReadMe.LLM. In contrast, Claude 3.7 already achieved 60\% with no context and 20\% with ReadMe.md, and likewise hits 100\% once provided with ReadMe.LLM. Similar to the point we made for DigitalRF's case study, the improvement from human-oriented docs to ReadMe.LLM is far more dramatic for Claude 3.5 than for Claude 3.7. 

\begin{figure}[H]
    \centering
    \includegraphics[width=0.6\textwidth]{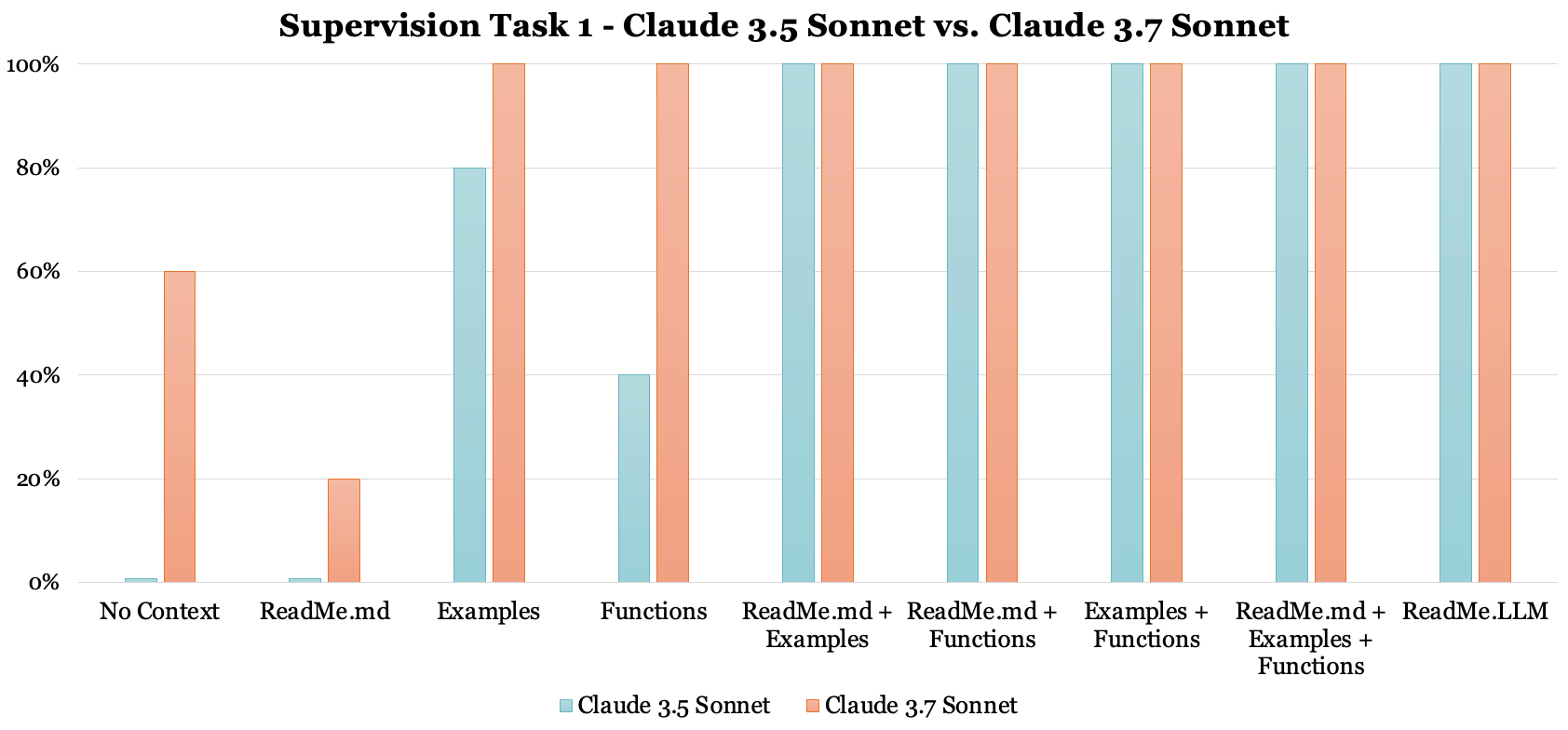}
    \caption{Supervision Task 1 - Claude 3.5 Sonnet vs. Claude 3.7 Sonnet}
\end{figure}

\section{Appendix: ReadMe.LLM Instructions}\label{app:readmellm_instructions}
\begin{enumerate}
    \item Determine Functions to Include: Select the most popular or important functions and classes that represent the library's core purpose and domain context.

    \item Extract Function Signatures: Clearly include function and method definitions with parameters and return types
    \item Find/Create Code Examples of Each Function: Pair signatures with illustrative code examples that demonstrate real-world usage of the functions
    \item Interleave Description of Function, Function Signature, and Example: Structure the content using sections with XML tags, combining a description, the function signatures, and the examples in a parsable layout. Below is the structure of the core building block of our ReadMe.LLM:
\begin{verbatim}
<context_1>
<context_1_description>
…
</context_1_description>

<context_1_function>
…
</context_1_function>

<context_1_example>
…
</context_1_example>
</context_1>
\end{verbatim}

\item Add rules to the top of the ReadMe.LLM file. Below are the rules we added to our ReadMe.LLM files:

\textbf{Rules:}
\begin{enumerate}
    \item \textbf{Rule number 1:} When you’re unsure about something, ask the user what information you need.
    \item \textbf{Rule number 2:} Reuse SuperVision’s functions and code when applicable.
    \item \textbf{Rule number 3:} Consider library dependencies when generating code solutions.
\end{enumerate}

\end{enumerate}

\clearpage

\section{Appendix: ReadMe.LLM Content}\label{app:readmellm_content}

\subsection{Supervision's ReadMe.LLM}\label{app:supervision_readmellm}

The following is the full ReadMe.LLM for the Supervision library:

\begin{lstlisting}

Rules:
Rule number 1: When you are unsure about something, ask the user what information you need.
Rule number 2: Reuse SuperVision functions and code when applicable
Rule number 3: Consider library dependencies when generating code solutions

<ReadMe.LLM>

<context_description>
The context will be for the Supervision Library. The Supervision library creates tools to enable developers to complete their Computer Vision tasks. The context I am giving you will be functions and examples related to detection, classification, and utilities from the Supervision Library. The context is organized into different numbered sections in order using XML tags. Within each section, there is a context description for that section, a code snippet, and use case examples.
</context_description>

<context_1>
<context_1_description>
The sv.Detections class in the Supervision library standardizes results from various object detection and segmentation models into a consistent format. This class simplifies data manipulation and filtering, providing a uniform API for integration with Supervision trackers, annotators, and tools.
</context_1_description>

<context_1_code_snippet>
@dataclass
class Detections:
    xyxy: np.ndarray
    mask: Optional[np.ndarray] = None
    confidence: Optional[np.ndarray] = None
    class_id: Optional[np.ndarray] = None
    tracker_id: Optional[np.ndarray] = None
    data: Dict[str, Union[np.ndarray, List]] = field(default_factory=dict)
    metadata: Dict[str, Any] = field(default_factory=dict)

    def __post_init__(self):
    
    def __len__(self):
        
    def __iter__(self) -> Iterator[Tuple[np.ndarray, Optional[np.ndarray], Optional[float], Optional[int], Optional[int], Dict[str, Union[np.ndarray, List]]]]:
        
    def __eq__(self, other: Detections):
        
    @classmethod
    def from_yolov5(cls, yolov5_results) -> Detections:
        
    @classmethod
    def from_ultralytics(cls, ultralytics_results) -> Detections:
        
    @classmethod
    def from_yolo_nas(cls, yolo_nas_results) -> Detections:
        
    @classmethod
    def from_tensorflow(cls, tensorflow_results: dict, resolution_wh: tuple) -> Detections:
        
    @classmethod
    def from_deepsparse(cls, deepsparse_results) -> Detections:
        
    @classmethod
    def from_mmdetection(cls, mmdet_results) -> Detections:
        
    @classmethod
    def from_transformers(cls, transformers_results: dict, id2label: Optional[Dict[int, str]] = None) -> Detections:
        
    @classmethod
    def from_detectron2(cls, detectron2_results: Any) -> Detections:
        
    @classmethod
    def from_inference(cls, roboflow_result: Union[dict, Any]) -> Detections:
        
    @classmethod
    def from_sam(cls, sam_result: List[dict]) -> Detections:
        
    @classmethod
    def from_azure_analyze_image(cls, azure_result: dict, class_map: Optional[Dict[int, str]] = None) -> Detections:
        
    @classmethod
    def from_paddledet(cls, paddledet_result) -> Detections:
        
    @classmethod
    def from_lmm(cls, lmm: Union[LMM, str], result: Union[str, dict], **kwargs: Any) -> Detections:
        
    @classmethod
    def from_vlm(cls, vlm: Union[VLM, str], result: Union[str, dict], **kwargs: Any) -> Detections:
        
    @classmethod
    def from_easyocr(cls, easyocr_results: list) -> Detections:
        
    @classmethod
    def from_ncnn(cls, ncnn_results) -> Detections:
        
    @classmethod
    def empty(cls) -> Detections:
        
    def is_empty(self) -> bool:
        
    @classmethod
    def merge(cls, detections_list: List[Detections]) -> Detections:
        
    def get_anchors_coordinates(self, anchor: Position) -> np.ndarray:
        
    def __getitem__(self, index: Union[int, slice, List[int], np.ndarray, str]) -> Union[Detections, List, np.ndarray, None]:
        
    def __setitem__(self, key: str, value: Union[np.ndarray, List]):
        
    @property
    def area(self) -> np.ndarray:
        
    @property
    def box_area(self) -> np.ndarray:
        
    def with_nms(self, threshold: float = 0.5, class_agnostic: bool = False) -> Detections:
        
    def with_nmm(self, threshold: float = 0.5, class_agnostic: bool = False) -> Detections:
        
def merge_inner_detection_object_pair(detections_1: Detections, detections_2: Detections) -> Detections:
    
def merge_inner_detections_objects(detections: List[Detections], threshold=0.5) -> Detections:
    
def validate_fields_both_defined_or_none(detections_1: Detections, detections_2: Detections) -> None:
</context_1_code_snippet>

<context_1_examples>
=== "Inference"
Use [`sv.Detections.from_inference`](/detection/core/#supervision.detection.core.Detections.from_inference) method, which accepts model results from both detection and segmentation models.

        ```python
        import cv2
        import supervision as sv
        from inference import get_model

        model = get_model(model_id="yolov8n-640")
        image = cv2.imread(<SOURCE_IMAGE_PATH>)
        results = model.infer(image)[0]
        detections = sv.Detections.from_inference(results)
        ```

=== "Ultralytics"
Use [`sv.Detections.from_ultralytics`](/detection/core/#supervision.detection.core.Detections.from_ultralytics) method, which accepts model results from both detection and segmentation models.

        ```python
        import cv2
        import supervision as sv
        from ultralytics import YOLO

        model = YOLO("yolov8n.pt")
        image = cv2.imread(<SOURCE_IMAGE_PATH>)
        results = model(image)[0]
        detections = sv.Detections.from_ultralytics(results)
        ```

=== "Transformers"
Use [`sv.Detections.from_transformers`](/detection/core/#supervision.detection.core.Detections.from_transformers) 
 method, which accepts model results from both detection and segmentation models.

        ```python
        import torch
        import supervision as sv
        from PIL import Image
        from transformers import DetrImageProcessor, DetrForObjectDetection

        processor = DetrImageProcessor.from_pretrained("facebook/detr-resnet-50")
        model = DetrForObjectDetection.from_pretrained("facebook/detr-resnet-50")

        image = Image.open(<SOURCE_IMAGE_PATH>)
        inputs = processor(images=image, return_tensors="pt")

        with torch.no_grad():
            outputs = model(**inputs)

        width, height = image.size
        target_size = torch.tensor([[height, width]])
        results = processor.post_process_object_detection(
            outputs=outputs, target_sizes=target_size)[0]
        detections = sv.Detections.from_transformers(
            transformers_results=results,
            id2label=model.config.id2label)
        ```
</context_1_examples>
</context_1>

<context_2>
<context_2_description>
This context section focuses on visual annotation utilities for object detection tasks from the SuperVision library. It provides various annotators to overlay bounding boxes, oriented boxes, masks, polygons, labels, and other graphical elements onto images based on object detection outputs.
</context_2_description>

<context_2_code_snippet>
class BoxAnnotator(BaseAnnotator):
    def __init__(self, color: Union[Color, ColorPalette], thickness: int, color_lookup: ColorLookup):
    def annotate(self, scene: ImageType, detections: Detections, custom_color_lookup: Optional[np.ndarray]) -> ImageType:

class OrientedBoxAnnotator(BaseAnnotator):
    def __init__(self, color: Union[Color, ColorPalette], thickness: int, color_lookup: ColorLookup):
    def annotate(self, scene: ImageType, detections: Detections, custom_color_lookup: Optional[np.ndarray]) -> ImageType:

class MaskAnnotator(BaseAnnotator):
    def __init__(self, color: Union[Color, ColorPalette], opacity: float, color_lookup: ColorLookup):
    def annotate(self, scene: ImageType, detections: Detections, custom_color_lookup: Optional[np.ndarray]) -> ImageType:

class PolygonAnnotator(BaseAnnotator):
    def __init__(self, color: Union[Color, ColorPalette], thickness: int, color_lookup: ColorLookup):
    def annotate(self, scene: ImageType, detections: Detections, custom_color_lookup: Optional[np.ndarray]) -> ImageType:

class ColorAnnotator(BaseAnnotator):
    def __init__(self, color: Union[Color, ColorPalette], opacity: float, color_lookup: ColorLookup):
    def annotate(self, scene: ImageType, detections: Detections, custom_color_lookup: Optional[np.ndarray]) -> ImageType:

class HaloAnnotator(BaseAnnotator):
    def __init__(self, color: Union[Color, ColorPalette], opacity: float, kernel_size: int, color_lookup: ColorLookup):
    def annotate(self, scene: ImageType, detections: Detections, custom_color_lookup: Optional[np.ndarray]) -> ImageType:

class EllipseAnnotator(BaseAnnotator):
    def __init__(self, color: Union[Color, ColorPalette], thickness: int, start_angle: int, end_angle: int, color_lookup: ColorLookup):
    def annotate(self, scene: ImageType, detections: Detections, custom_color_lookup: Optional[np.ndarray]) -> ImageType:

class BoxCornerAnnotator(BaseAnnotator):
    def __init__(self, color: Union[Color, ColorPalette], thickness: int, corner_length: int, color_lookup: ColorLookup):
    def annotate(self, scene: ImageType, detections: Detections, custom_color_lookup: Optional[np.ndarray]) -> ImageType:

class CircleAnnotator(BaseAnnotator):
    def __init__(self, color: Union[Color, ColorPalette], thickness: int, color_lookup: ColorLookup):
    def annotate(self, scene: ImageType, detections: Detections, custom_color_lookup: Optional[np.ndarray]) -> ImageType:

class DotAnnotator(BaseAnnotator):
    def __init__(self, color: Union[Color, ColorPalette], radius: int, position: Position, color_lookup: ColorLookup, outline_thickness: int, outline_color: Union[Color, ColorPalette]):
    def annotate(self, scene: ImageType, detections: Detections, custom_color_lookup: Optional[np.ndarray]) -> ImageType:

class LabelAnnotator(BaseAnnotator):
    def __init__(self, color: Union[Color, ColorPalette], text_color: Union[Color, ColorPalette], text_scale: float, text_thickness: int, text_padding: int, text_position: Position, color_lookup: ColorLookup, border_radius: int, smart_position: bool):
    def annotate(self, scene: ImageType, detections: Detections, labels: Optional[List[str]], custom_color_lookup: Optional[np.ndarray]) -> ImageType:
    def _validate_labels(self, labels: Optional[List[str]], detections: Detections):
    def _get_label_properties(self, detections: Detections, labels: List[str]) -> np.ndarray:
    def _get_labels_text(self, detections: Detections, custom_labels: Optional[List[str]]) -> List[str]:
    def _draw_labels(self, scene: np.ndarray, labels: List[str], label_properties: np.ndarray, detections: Detections, custom_color_lookup: Optional[np.ndarray]) -> None:
    def draw_rounded_rectangle(self, scene: np.ndarray, xyxy: Tuple[int, int, int, int], color: Tuple[int, int, int], border_radius: int) -> np.ndarray:

class RichLabelAnnotator(BaseAnnotator):
    def __init__(self, color: Union[Color, ColorPalette], text_color: Union[Color, ColorPalette], font_path: Optional[str], font_size: int, text_padding: int, text_position: Position, color_lookup: ColorLookup, border_radius: int, smart_position: bool):
    def annotate(self, scene: ImageType, detections: Detections, labels: Optional[List[str]], custom_color_lookup: Optional[np.ndarray]) -> ImageType:
    def _validate_labels(self, labels: Optional[List[str]], detections: Detections):
    def _get_label_properties(self, draw, detections: Detections, labels: List[str]) -> np.ndarray:
    def _get_labels_text(self, detections: Detections, custom_labels: Optional[List[str]]) -> List[str]:
    def _draw_labels(self, draw, labels: List[str], label_properties: np.ndarray, detections: Detections, custom_color_lookup: Optional[np.ndarray]) -> None:
    def _load_font(self, font_size: int, font_path: Optional[str]):

class IconAnnotator(BaseAnnotator):
    def __init__(self, icon_resolution_wh: Tuple[int, int] = (64, 64), icon_position: Position = Position.TOP_CENTER, offset_xy: Tuple[int, int] = (0, 0)):
    @ensure_cv2_image_for_annotation
    def annotate(self, scene: ImageType, detections: Detections, icon_path: Union[str, List[str]]) -> ImageType:
    @lru_cache
    def _load_icon(self, icon_path: str) -> np.ndarray:

class BlurAnnotator(BaseAnnotator):
    def __init__(self, kernel_size: int = 15):
    @ensure_cv2_image_for_annotation
    def annotate(self, scene: ImageType, detections: Detections) -> ImageType:

class TraceAnnotator(BaseAnnotator):
    def __init__(self, color: Union[Color, ColorPalette] = ColorPalette.DEFAULT, position: Position = Position.CENTER, trace_length: int = 30, thickness: int = 2, color_lookup: ColorLookup = ColorLookup.CLASS):
    @ensure_cv2_image_for_annotation
    def annotate(self, scene: ImageType, detections: Detections, custom_color_lookup: Optional[np.ndarray] = None) -> ImageType:

class HeatMapAnnotator(BaseAnnotator):
    def __init__(self, position: Position = Position.BOTTOM_CENTER, opacity: float = 0.2, radius: int = 40, kernel_size: int = 25, top_hue: int = 0, low_hue: int = 125):
    @ensure_cv2_image_for_annotation
    def annotate(self, scene: ImageType, detections: Detections) -> ImageType:

class PixelateAnnotator(BaseAnnotator):
    def __init__(self, pixel_size: int = 20):
    @ensure_cv2_image_for_annotation
    def annotate(self, scene: ImageType, detections: Detections) -> ImageType:

class TriangleAnnotator(BaseAnnotator):
    def __init__(self, color: Union[Color, ColorPalette] = ColorPalette.DEFAULT, base: int = 10, height: int = 10, position: Position = Position.TOP_CENTER, color_lookup: ColorLookup = ColorLookup.CLASS, outline_thickness: int = 0, outline_color: Union[Color, ColorPalette] = Color.BLACK):
    @ensure_cv2_image_for_annotation
    def annotate(self, scene: ImageType, detections: Detections, custom_color_lookup: Optional[np.ndarray] = None) -> ImageType:

class RoundBoxAnnotator(BaseAnnotator):
    def __init__(self, color: Union[Color, ColorPalette] = ColorPalette.DEFAULT, thickness: int = 2, color_lookup: ColorLookup = ColorLookup.CLASS, roundness: float = 0.6):
    @ensure_cv2_image_for_annotation
    def annotate(self, scene: ImageType, detections: Detections, custom_color_lookup: Optional[np.ndarray] = None) -> ImageType:

class PercentageBarAnnotator(BaseAnnotator):
    def __init__(self, height: int = 16, width: int = 80, color: Union[Color, ColorPalette] = ColorPalette.DEFAULT, border_color: Color = Color.BLACK, position: Position = Position.TOP_CENTER, color_lookup: ColorLookup = ColorLookup.CLASS, border_thickness: Optional[int] = None):
    @ensure_cv2_image_for_annotation
    def annotate(self, scene: ImageType, detections: Detections, custom_color_lookup: Optional[np.ndarray] = None, custom_values: Optional[np.ndarray] = None) -> ImageType:
    @staticmethod
    def calculate_border_coordinates(anchor_xy: Tuple[int, int], border_wh: Tuple[int, int], position: Position) -> Tuple[Tuple[int, int], Tuple[int, int]]:
    @staticmethod
    def validate_custom_values(custom_values: Optional[Union[np.ndarray, List[float]]], detections: Detections) -> None:

class CropAnnotator(BaseAnnotator):
    def __init__(self, position: Position = Position.TOP_CENTER, scale_factor: float = 2.0, border_color: Union[Color, ColorPalette] = ColorPalette.DEFAULT, border_thickness: int = 2, border_color_lookup: ColorLookup = ColorLookup.CLASS):
    @ensure_cv2_image_for_annotation
    def annotate(self, scene: ImageType, detections: Detections, custom_color_lookup: Optional[np.ndarray] = None) -> ImageType:
    @staticmethod
    def calculate_crop_coordinates(anchor: Tuple[int, int], crop_wh: Tuple[int, int], position: Position) -> Tuple[Tuple[int, int], Tuple[int, int]]:

class BackgroundOverlayAnnotator(BaseAnnotator):
    def __init__(self, color: Color = Color.BLACK, opacity: float = 0.5, force_box: bool = False):
    @ensure_cv2_image_for_annotation
    def annotate(self, scene: ImageType, detections: Detections) -> ImageType:

class ComparisonAnnotator:
    def __init__(self, color_1: Color = Color.RED, color_2: Color = Color.GREEN, color_overlap: Color = Color.BLUE, *, opacity: float = 0.75, label_1: str = "", label_2: str = "", label_overlap: str = "", label_scale: float = 1.0):
    @ensure_cv2_image_for_annotation
    def annotate(self, scene: ImageType, detections_1: Detections, detections_2: Detections) -> ImageType:
    @staticmethod
    def _use_obb(detections_1: Detections, detections_2: Detections) -> bool:
    @staticmethod
    def _use_mask(detections_1: Detections, detections_2: Detections) -> bool:
    @staticmethod
    def _mask_from_xyxy(scene: np.ndarray, detections: Detections) -> np.ndarray:
    @staticmethod
    def _mask_from_obb(scene: np.ndarray, detections: Detections) -> np.ndarray:
    @staticmethod
    def _mask_from_mask(scene: np.ndarray, detections: Detections) -> np.ndarray:
    def _draw_labels(self, scene: np.ndarray) -> None:
</context_2_code_snippet>

<context_2_examples	>
image = ...
detections = sv.Detections(...)

# Box Annotator Example
box_annotator = sv.BoxAnnotator()
annotated_frame = box_annotator.annotate(scene=image.copy(), detections=detections)

# Box Corner Annotator Example
corner_annotator = sv.BoxCornerAnnotator()
annotated_frame = corner_annotator.annotate(scene=image.copy(), detections=detections)

# Color Annotator Example
color_annotator = sv.ColorAnnotator()
annotated_frame = color_annotator.annotate(scene=image.copy(), detections=detections)

# Circle Annotator Example
circle_annotator = sv.CircleAnnotator()
annotated_frame = circle_annotator.annotate(scene=image.copy(), detections=detections)

# Dot Annotator Example
dot_annotator = sv.DotAnnotator()
annotated_frame = dot_annotator.annotate(scene=image.copy(), detections=detections)

# Triangle Annotator Example
triangle_annotator = sv.TriangleAnnotator()
annotated_frame = triangle_annotator.annotate(scene=image.copy(), detections=detections)

# Ellipse Annotator Example
ellipse_annotator = sv.EllipseAnnotator()
annotated_frame = ellipse_annotator.annotate(scene=image.copy(), detections=detections)

# Halo Annotator Example
halo_annotator = sv.HaloAnnotator()
annotated_frame = halo_annotator.annotate(scene=image.copy(), detections=detections)

# Percentage Bar Annotator Example
percentage_bar_annotator = sv.PercentageBarAnnotator()
annotated_frame = percentage_bar_annotator.annotate(scene=image.copy(), detections=detections)

# Mask Annotator Example
mask_annotator = sv.MaskAnnotator()
annotated_frame = mask_annotator.annotate(scene=image.copy(), detections=detections)

# Polygon Annotator Example
polygon_annotator = sv.PolygonAnnotator()
annotated_frame = polygon_annotator.annotate(scene=image.copy(), detections=detections)

# Label Annotator Example
labels = [f"{class_name} {confidence:.2f}" for class_name, confidence in zip(detections['class_name'], detections.confidence)]
label_annotator = sv.LabelAnnotator(text_position=sv.Position.CENTER)
annotated_frame = label_annotator.annotate(scene=image.copy(), detections=detections, labels=labels)

# Rich Label Annotator Example
labels = [f"{class_name} {confidence:.2f}" for class_name, confidence in zip(detections['class_name'], detections.confidence)]
rich_label_annotator = sv.RichLabelAnnotator(font_path="<TTF_FONT_PATH>", text_position=sv.Position.CENTER)
annotated_frame = rich_label_annotator.annotate(scene=image.copy(), detections=detections, labels=labels)

# Icon Annotator Example
icon_paths = ["<ICON_PATH>" for _ in detections]
icon_annotator = sv.IconAnnotator()
annotated_frame = icon_annotator.annotate(scene=image.copy(), detections=detections, icon_path=icon_paths)

# Blur Annotator Example
blur_annotator = sv.BlurAnnotator()
annotated_frame = blur_annotator.annotate(scene=image.copy(), detections=detections)

# Pixelate Annotator Example
pixelate_annotator = sv.PixelateAnnotator()
annotated_frame = pixelate_annotator.annotate(scene=image.copy(), detections=detections)

# Trace Annotator Example
model = YOLO('yolov8x.pt')
trace_annotator = sv.TraceAnnotator()
video_info = sv.VideoInfo.from_video_path(video_path='...')
frames_generator = sv.get_video_frames_generator(source_path='...')
tracker = sv.ByteTrack()

with sv.VideoSink(target_path='...', video_info=video_info) as sink:
    for frame in frames_generator:
        result = model(frame)[0]
        detections = sv.Detections.from_ultralytics(result)
        detections = tracker.update_with_detections(detections)
        annotated_frame = trace_annotator.annotate(scene=frame.copy(), detections=detections)
        sink.write_frame(frame=annotated_frame)

# Heat Map Annotator Example
model = YOLO('yolov8x.pt')
heat_map_annotator = sv.HeatMapAnnotator()
video_info = sv.VideoInfo.from_video_path(video_path='...')
frames_generator = sv.get_video_frames_generator(source_path='...')

with sv.VideoSink(target_path='...', video_info=video_info) as sink:
    for frame in frames_generator:
        result = model(frame)[0]
        detections = sv.Detections.from_ultralytics(result)
        annotated_frame = heat_map_annotator.annotate(scene=frame.copy(), detections=detections)
        sink.write_frame(frame=annotated_frame)

# Background Overlay Annotator Example
background_overlay_annotator = sv.BackgroundOverlayAnnotator()
annotated_frame = background_overlay_annotator.annotate(scene=image.copy(), detections=detections)

# Comparison Annotator Example
image = ...
detections_1 = sv.Detections(...)
detections_2 = sv.Detections(...)
comparison_annotator = sv.ComparisonAnnotator()
annotated_frame = comparison_annotator.annotate(scene=image.copy(), detections_1=detections_1, detections_2=detections_2)
</context_2_examples>
</context_2>

<context_3>
<context_3_description>
The Supervision library provides a set of utilities for image preprocessing, overlaying annotations, creating image grids, and saving image outputs.
</context_3_description>

<context_3_code_snippet>
def crop_image(
    image: ImageType,
    xyxy: Union[npt.NDArray[int], List[int], Tuple[int, int, int, int]],
) -> ImageType:

def scale_image(image: ImageType, scale_factor: float) -> ImageType:

def resize_image(
    image: ImageType,
    resolution_wh: Tuple[int, int],
    keep_aspect_ratio: bool = False,
) -> ImageType:

def letterbox_image(
    image: ImageType,
    resolution_wh: Tuple[int, int],
    color: Union[Tuple[int, int, int], Color] = Color.BLACK,
) -> ImageType:

def overlay_image(
    image: npt.NDArray[np.uint8],
    overlay: npt.NDArray[np.uint8],
    anchor: Tuple[int, int],
) -> npt.NDArray[np.uint8]:

class ImageSink:
    def __init__(
        self,
        target_dir_path: str,
        overwrite: bool = False,
        image_name_pattern: str = "image_{:05d}.png",
    ):

    def __enter__(self):

    def save_image(self, image: np.ndarray, image_name: Optional[str] = None):

    def __exit__(self, exc_type, exc_value, exc_traceback):

def create_tiles(
    images: List[ImageType],
    grid_size: Optional[Tuple[Optional[int], Optional[int]]] = None,
    single_tile_size: Optional[Tuple[int, int]] = None,
    tile_scaling: Literal["min", "max", "avg"] = "avg",
    tile_padding_color: Union[Tuple[int, int, int], Color] = Color.from_hex("#D9D9D9"),
    tile_margin: int = 10,
    tile_margin_color: Union[Tuple[int, int, int], Color] = Color.from_hex("#BFBEBD"),
    return_type: Literal["auto", "cv2", "pillow"] = "auto",
    titles: Optional[List[Optional[str]]] = None,
    titles_anchors: Optional[Union[Point, List[Optional[Point]]]] = None,
    titles_color: Union[Tuple[int, int, int], Color] = Color.from_hex("#262523"),
    titles_scale: Optional[float] = None,
    titles_thickness: int = 1,
    titles_padding: int = 10,
    titles_text_font: int = cv2.FONT_HERSHEY_SIMPLEX,
    titles_background_color: Union[Tuple[int, int, int], Color] = Color.from_hex(
        "#D9D9D9"
    ),
    default_title_placement: RelativePosition = "top",
) -> ImageType:

def _negotiate_tiles_format(images: List[ImageType]) -> Literal["cv2", "pillow"]:

def _calculate_aggregated_images_shape(
    images: List[np.ndarray], aggregator: Callable[[List[int]], float]
) -> Tuple[int, int]:

def _aggregate_images_shape(
    images: List[np.ndarray], mode: Literal["min", "max", "avg"]
) -> Tuple[int, int]:

def _establish_grid_size(
    images: List[np.ndarray], grid_size: Optional[Tuple[Optional[int], Optional[int]]]
) -> Tuple[int, int]:

def _negotiate_grid_size(images: List[np.ndarray]) -> Tuple[int, int]:

def _generate_tiles(
    images: List[np.ndarray],
    grid_size: Tuple[int, int],
    single_tile_size: Tuple[int, int],
    tile_padding_color: Tuple[int, int, int],
    tile_margin: int,
    tile_margin_color: Tuple[int, int, int],
    titles: Optional[List[Optional[str]]],
    titles_anchors: List[Optional[Point]],
    titles_color: Tuple[int, int, int],
    titles_scale: Optional[float],
    titles_thickness: int,
    titles_padding: int,
    titles_text_font: int,
    titles_background_color: Tuple[int, int, int],
    default_title_placement: RelativePosition,
) -> np.ndarray:

def _draw_texts(
    images: List[np.ndarray],
    titles: Optional[List[Optional[str]]],
    titles_anchors: List[Optional[Point]],
    titles_color: Tuple[int, int, int],
    titles_scale: Optional[float],
    titles_thickness: int,
    titles_padding: int,
    titles_text_font: int,
    titles_background_color: Tuple[int, int, int],
    default_title_placement: RelativePosition,
) -> List[np.ndarray]:

def _prepare_default_titles_anchors(
    images: List[np.ndarray],
    titles_anchors: List[Optional[Point]],
    default_title_placement: RelativePosition,
) -> List[Point]:

def _merge_tiles_elements(
    tiles_elements: List[List[np.ndarray]],
    grid_size: Tuple[int, int],
    single_tile_size: Tuple[int, int],
    tile_margin: int,
    tile_margin_color: Tuple[int, int, int],
) -> np.ndarray:

def _generate_color_image(
    shape: Tuple[int, int], color: Tuple[int, int, int]
) -> np.ndarray:
</context_3_code_snippet>

<context_3_examples>
# Using sv.crop_image to crop an image based on bounding box coordinates
image = cv2.imread(<SOURCE_IMAGE_PATH>)
image.shape
# (1080, 1920, 3)

xyxy = [200, 400, 600, 800]
cropped_image = sv.crop_image(image=image, xyxy=xyxy)
cropped_image.shape
# (400, 400, 3)

# Using sv.scale_image to scale an image by a given factor
scaled_image = sv.scale_image(image=image, scale_factor=0.5)
scaled_image.shape
# (540, 960, 3)

# Using sv.resize_image to resize an image while optionally maintaining aspect ratio
resized_image = sv.resize_image(
    image=image, resolution_wh=(1000, 1000), keep_aspect_ratio=True
)
resized_image.shape
# (562, 1000, 3)

# Using sv.letterbox_image to resize an image while maintaining aspect ratio and adding padding
letterboxed_image = sv.letterbox_image(image=image, resolution_wh=(1000, 1000))
letterboxed_image.shape
# (1000, 1000, 3)

# Using sv.overlay_image to overlay an image onto another at a specified anchor point
image = cv2.imread(<SOURCE_IMAGE_PATH>)
overlay = np.zeros((400, 400, 3), dtype=np.uint8)
result_image = sv.overlay_image(image=image, overlay=overlay, anchor=(200, 400))

# Using sv.ImageSink to save images in a specified directory
frames_generator = sv.get_video_frames_generator(<SOURCE_VIDEO_PATH>, stride=2)

with sv.ImageSink(target_dir_path=<TARGET_CROPS_DIRECTORY>) as sink:
    for image in frames_generator:
        sink.save_image(image=image)

</context_3_examples>
</context_3>
</ReadMe.LLM>
\end{lstlisting}

\clearpage

\subsection{DigitalRF's ReadMe.LLM}\label{app:digitalRF_readmellm}

The following is the full ReadMe.LLM for the DigitalRF library:

\begin{lstlisting}
<ReadMe.LLM>

<context_description>
The context will be for the DigitalRF library. The Digital RF project encompasses a standardized HDF5 format for reading and writing of radio frequency data and the software for doing so. The format is designed to be self-documenting for data archive and to allow rapid random access for data processing. The context I am giving you will be functions and examples from the DigitalRF library. The context is organized into different numbered sections in order using XML tags. Within each section, there is a context description for that section, a code snippet, and use case examples.
</context_description>

<context_1>
<context_1_description>
The DigitalRFWriter class is responsible for writing RF data in the Digital RF HDF5 format. It organizes data into time-stamped subdirectories and files based on configurable cadences and ensures data integrity through type casting, optional compression, and checksum functionality.
</context_1_description>

<context_1_code_snippet>
class DigitalRFWriter(object):
    """Write a channel of data in Digital RF HDF5 format."""

    _writer_version = libdigital_rf_version

    def __init__(
        self,
        directory,
        dtype,
        subdir_cadence_secs,
        file_cadence_millisecs,
        start_global_index,
        sample_rate_numerator,
        sample_rate_denominator,
        uuid_str=None,
        compression_level=0,
        checksum=False,
        is_complex=True,
        num_subchannels=1,
        is_continuous=True,
        marching_periods=True,
    ):
        """Initialize writer to channel directory with given parameters.

        Parameters
        ----------
        directory : string
            The directory where this channel is to be written. It must already
            exist and be writable.

        dtype : np.dtype | object to be cast by np.dtype()
            Object that gives the numpy dtype of the data to be written. This
            value is passed into ``np.dtype`` to get the actual dtype
            (e.g. ``np.dtype('>i4')``). Scalar types, complex types, and
            structured complex types with 'r' and 'i' fields of scalar types
            are valid.

        subdir_cadence_secs : int
            The number of seconds of data to store in one subdirectory. The
            timestamp of any subdirectory will be an integer multiple of this
            value.

        file_cadence_millisecs : int
            The number of milliseconds of data to store in each file. Note that
            an integer number of files must exactly span a subdirectory,
            implying::

                (subdir_cadence_secs*1000 % file_cadence_millisecs) == 0

        start_global_index : int
            The index of the first sample given in number of samples since the
            epoch. For a given ``start_time`` in seconds since the epoch, this
            can be calculated as::

                floor(start_time * (np.longdouble(sample_rate_numerator) /
                                    np.longdouble(sample_rate_denominator)))

        sample_rate_numerator : int
            Numerator of sample rate in Hz.

        sample_rate_denominator : int
            Denominator of sample rate in Hz.

        Other Parameters
        ----------------
        uuid_str : None | string, optional
            UUID string that will act as a unique identifier for the data and
            can be used to tie the data files to metadata. If None, a random
            UUID will be generated.

        compression_level : int, optional
            0 for no compression (default), 1-9 for varying levels of gzip
            compression (1 == least compression, least CPU; 9 == most
            compression, most CPU).

        checksum : bool, optional
            If True, use Hdf5 checksum capability. If False (default), no
            checksum.

        is_complex : bool, optional
            This parameter is only used when `dtype` is not complex.
            If True (the default), interpret supplied data as interleaved
            complex I/Q samples. If False, each sample has a single value.

        num_subchannels : int, optional
            Number of subchannels to write simultaneously. Default is 1.

        is_continuous : bool, optional
            If True, data will be written in continuous blocks. If False data
            will be written with gapped blocks. Fastest write/read speed is
            achieved with `is_continuous` True, `checksum` False, and
            `compression_level` 0 (all defaults).

        marching_periods : bool, optional
            If True, write a period to stdout for every file when
            writing.
        """

    def __enter__(self):
        """Enter method to enable context manager `with` statement."""

    def __exit__(self, exc_type, exc_value, traceback):
        """Exit method to enable context manager `with` statement."""

    @classmethod
    def get_version(cls):
        """Return the version string of the Digital RF writer."""

    def rf_write(self, arr, next_sample=None):
        """Write the next in-sequence samples from a given array.

        Parameters
        ----------
        arr : array_like
            Array of data to write. The array must have the same number of
            subchannels and type as declared when initializing the writer
            object, or an error will be raised. For single valued data, number
            of columns == number of subchannels. For complex data, there are
            two sizes of input arrays that are allowed:
                1. For a complex array or a structured array with column names
                    'r' and 'i' (as stored in the HDF5 file), the shape must
                    be (N, num_subchannels).
                2. For a non-structured, non-complex array, the shape must be
                    (N, 2*num_subchannels). I/Q are assumed to be interleaved.

        next_sample : long, optional
            Index of next sample to write relative to `start_global_index` of
            the first sample. If None (default), the array will be written to
            the next available sample after previous writes,
            `self._next_avail_sample`. A ValueError is raised if `next_sample`
            is less than `self._next_avail_sample`.

        Returns
        -------
        next_avail_sample : int
            Index of the next available sample after the array has been
            written.

        See Also
        --------
        rf_write_blocks

        Notes
        -----
        Here's an example of one way to create a structured numpy array with
        complex data with dtype int16::

            arr_data = np.ones(
                (num_rows, num_subchannels),
                dtype=[('r', np.int16), ('i', np.int16)],
            )
            for i in range(num_subchannels):
                for j in range(num_rows):
                    arr_data[j,i]['r'] = 2
                    arr_data[j,i]['i'] = 3

        The same data could be also be passed as an interleaved array::

            arr_data = np.ones(
                (num_rows, num_subchannels*2),
                dtype=np.int16,
            )

        """

    def rf_write_blocks(self, arr, global_sample_arr, block_sample_arr):
        """Write blocks of data with interleaved gaps.

        Parameters
        ----------
        arr : array_like
            Array of data to write. See `rf_write` for a complete description
            of allowed forms.

        global_sample_arr : array_like of shape (N,) and type uint64
            An array that sets the global sample index (relative to
            `start_global_index` of the first sample) for each continuous
            block of data in arr. The values must be increasing, and the first
            value must be >= self._next_avail_sample or a ValueError raised.

        block_sample_arr : array_like of shape (N,) and type uint64
            An array that gives the index into arr for the start of each
            continuous block. The first value must be 0, and all values
            must be < len(arr). Increments between values must be > 0 and less
            than the corresponding increment in `global_sample_arr`.

        Returns
        -------
        next_avail_sample : int
            Index of the next available sample after the array has been
            written.


        See Also
        --------
        rf_write

        """

    def get_total_samples_written(self):
        """Return the total number of samples written in per channel.
        This does not include gaps.
        """
        return self._total_samples_written

    def get_next_available_sample(self):
        """Return the index of the next sample available for writing.
        This is equal to (total_samples_written + total_gap_samples).
        """
        return self._next_avail_sample

    def get_total_gap_samples(self):
        """Return the total number of samples contained in data gaps."""
        return self._total_gap_samples

    def get_last_file_written(self):
        """Return the full path to the last file written."""
        try:
            return _py_rf_write_hdf5.get_last_file_written(self._channelObj)
        except AttributeError:
            return self._last_file_written

    def get_last_dir_written(self):
        """Return the full path to the last directory written."""
        try:
            return _py_rf_write_hdf5.get_last_dir_written(self._channelObj)
        except AttributeError:
            return self._last_dir_written

    def get_last_utc_timestamp(self):
        """Return UTC timestamp of the time of the last data written."""
        try:
            return _py_rf_write_hdf5.get_last_utc_timestamp(self._channelObj)
        except AttributeError:
            return self._last_utc_timestamp

    def close(self):
        """Free memory of the underlying C object and close the last HDF5 file.
        No more data can be written using this writer instance after close has
        been called.
        """

    def _cast_input_array(self, arr):
        """Cast input array to correct type and check for the correct shape.

        Parameters
        ----------
        arr : array_like
            See `rf_write` method for a complete description of allowed values.

        Returns
        -------
        arr : ndarray of type self.dtype or self.structdtype

        Raises
        ------
        TypeError
            If the array type cannot be cast to the writer type.
        ValueError
            If the array shape does not match the specified number of
            subchannels.
        """

    def _cast_sample_array(self, sample_arr):
        """Cast sample array to equivalent values of uint64.

        Parameters
        ----------
        sample_arr : array_like
            Array of (global, block) sample indices.

        Returns
        -------
        sample_arr : ndarray of type uint64

        Raises
        ------
        TypeError
            If the array type cannot be cast to equivalent values of uint64.
        ValueError
            If the array is not 1-D.
        """
</context_1_code_snippet>

<context_1_examples>
"""A simple example of writing Digital RF with python.

Writes continuous complex short data.

"""
from __future__ import absolute_import, division, print_function

import os
import shutil
import tempfile

import digital_rf
import numpy as np

datadir = os.path.join(tempfile.gettempdir(), "example_digital_rf")
chdir = os.path.join(datadir, "junk0")

# writing parameters
sample_rate_numerator = int(100)  # 100 Hz sample rate - typically MUCH faster
sample_rate_denominator = 1
sample_rate = np.longdouble(sample_rate_numerator) / sample_rate_denominator
dtype_str = "i2"  # short int
sub_cadence_secs = (
    4  # Number of seconds of data in a subdirectory - typically MUCH larger
)
file_cadence_millisecs = 400  # Each file will have up to 400 ms of data
compression_level = 1  # low level of compression
checksum = False  # no checksum
is_complex = True  # complex values
is_continuous = True
num_subchannels = 1  # only one subchannel
marching_periods = False  # no marching periods when writing
uuid = "Fake UUID - use a better one!"
vector_length = 100  # number of samples written for each call - typically MUCH longer

# create short data in r/i to test using that to write
arr_data = np.ones(
    (vector_length, num_subchannels), dtype=[("r", np.int16), ("i", np.int16)]
)
for i in range(len(arr_data)):
    arr_data[i]["r"] = 2 * i
    arr_data[i]["i"] = 3 * i

# start 2014-03-09 12:30:30 plus one sample
start_global_index = int(np.uint64(1394368230 * sample_rate)) + 1

# set up top level directory
shutil.rmtree(chdir, ignore_errors=True)
os.makedirs(chdir)

print(
    (
        "Writing complex short to multiple files and subdirectores in {0}"
        " channel junk0"
    ).format(datadir)
)

# init
dwo = digital_rf.DigitalRFWriter(
    chdir,
    dtype_str,
    sub_cadence_secs,
    file_cadence_millisecs,
    start_global_index,
    sample_rate_numerator,
    sample_rate_denominator,
    uuid,
    compression_level,
    checksum,
    is_complex,
    num_subchannels,
    is_continuous,
    marching_periods,
)
# write
for i in range(7):  # will write 700 samples - so creates two subdirectories
    result = dwo.rf_write(arr_data)
    print("Last file written = %s" % (dwo.get_last_file_written()))
    print("Last dir written = %s" % (dwo.get_last_dir_written()))
    print("UTC timestamp of last write is %i" % (dwo.get_last_utc_timestamp()))

# close
dwo.close()
print("done test")
</context_1_examples>
</context_1>

<context_2>
<context_2_description>
The DigitalRFReader class is designed for random-access reading of RF data stored in the Digital RF HDF5 format. It allows users to retrieve continuous blocks of data, read metadata, and obtain vectorized data in various formats. With robust methods to handle file segmentation, gap detection, and channel-based organization, it provides a flexible interface for accessing and analyzing the stored RF data.
</context_2_description>

<context_2_code_snippet>
class DigitalRFReader(object):
    """Read data in Digital RF HDF5 format.
    This class allows random access to the rf data.
    """

    def __init__(self, top_level_directory_arg, rdcc_nbytes=4000000):
        """Initialize reader to directory containing Digital RF channels.

        Parameters
        ----------
        top_level_directory_arg : string
            Either a single top level directory containing Digital RF channel
            directories, or a list of such. A directory can be a file system
            path or a url, where the url points to a top level directory. Each
            must be a local path, or start with 'http://'', 'file://'', or
            'ftp://''.

        rdcc_nbytes : int, optional
            HDF5 chunk cache size. Needs to be at least file size
            for efficient access of compressed or checksummed files to
            avoid serious performance penalties with sparse data access.

        Notes
        -----
        A top level directory must contain files in the format:
            [channel]/[YYYY-MM-DDTHH-MM-SS]/rf@[seconds].[%03i milliseconds].h5

        If more than one top level directory contains the same channel_name
        subdirectory, this is considered the same channel. An error is raised
        if their sample rates differ, or if their time periods overlap.

        """

    def __enter__(self):
        """Enter method to enable context manager `with` statement."""

    def __exit__(self, exc_type, exc_value, traceback):
        """Exit method to enable context manager `with` statement."""

    def close(self):
        """Close reader object and any open associated HDF5 files.
        This object cannot be used once it is closed.
        """

    def get_channels(self):
        """Return an alphabetically sorted list of channels."""

    def read(self, start_sample, end_sample, channel_name, sub_channel=None):
        """Read continuous blocks of data between start and end samples.

        This is the basic read method, upon which more specialized read methods
        are based. For general use, `read_vector` is recommended. This method
        returns data as it is stored in the HDF5 file: in blocks of continous
        samples and with HDF5-native types (e.g. complex integer-typed data has
        a stuctured dtype with 'r' and 'i' fields).

        Parameters
        ----------
        start_sample : int
            Sample index for start of read, given in the number of samples
            since the epoch (time_since_epoch*sample_rate).

        end_sample : int
            Sample index for end of read (inclusive), given in the number of
            samples since the epoch (time_since_epoch*sample_rate).

        channel_name : string
            Name of channel to read from, one of ``get_channels()``.

        sub_channel : None | int, optional
            If None, the return array will contain all subchannels of data and
            be 2-d. If an integer, the return array will be 1-d and contain the
            data of the subchannel given by that integer index.

        Returns
        -------
        OrderedDict
            The dictionary's keys are the start sample of each continuous block
            found between `start_sample` and `end_sample` (inclusive). Each
            value is the numpy array of continous data starting at the key's
            index. The returned array has the same type as the data stored in
            the HDF5 file's rf_data dataset.

        See Also
        --------
        get_continuous_blocks : Similar, except no data is read.
        read_vector : Read data into a vector of floating-point type.
        read_vector_1d : Read data into a 1-d vector of floating-point type.
        read_vector_raw : Read data into a vector of HDF5-native type.

        """

    def get_bounds(self, channel_name):
        """Get indices of first- and last-known sample for a given channel.
        Parameters
        ----------
        channel_name : string
            Name of channel, one of ``get_channels()``.

        Returns
        -------
        first_sample_index : int | None
            Index of the first sample, given in the number of samples since the
            epoch (time_since_epoch*sample_rate).

        last_sample_index : int | None
            Index of the last sample, given in the number of samples since the
            epoch (time_since_epoch*sample_rate).

        """
        
    def get_properties(self, channel_name, sample=None):
        """Get dictionary of the properties particular to a Digital RF channel.

        Parameters
        ----------
        channel_name : string
            Name of channel, one of ``get_channels()``.

        sample : None | int
            If None, return the properties of the top-level drf_properties.h5
            file in the channel directory which applies to all samples. If a
            sample index is given, then return the properties particular to the
            file containing that sample index. This includes the top-level
            properties and additional attributes that can vary from file to
            file. If no data file is found associated with the input sample,
            then an IOError is raised.

        Returns
        -------
        dict
            Dictionary providing the properties.

        Notes
        -----
        The top-level properties, always returned, are:

            H5Tget_class : int
                Result of H5Tget_class(hdf5_data_object->hdf5_data_object)
            H5Tget_offset : int
                Result of H5Tget_offset(hdf5_data_object->hdf5_data_object)
            H5Tget_order : int
                Result of H5Tget_order(hdf5_data_object->hdf5_data_object)
            H5Tget_precision : int
                Result of H5Tget_precision(hdf5_data_object->hdf5_data_object)
            H5Tget_size : int
                Result of H5Tget_size(hdf5_data_object->hdf5_data_object)
            digital_rf_time_description : string
                Text description of Digital RF time conventions.
            digital_rf_version : string
                Version string of Digital RF writer.
            epoch : string
                Start time at sample 0 (always 1970-01-01 UT midnight)
            file_cadence_millisecs : int
            is_complex : int
            is_continuous : int
            num_subchannels : int
            sample_rate_numerator : int
            sample_rate_denominator : int
            samples_per_second : np.longdouble
            subdir_cadence_secs : int

        The additional properties particular to each file are:

            computer_time : int
                Unix time of initial file creation.
            init_utc_timestamp : int
                Changes at each restart of the recorder - needed if leap
                seconds correction applied.
            sequence_num : int
                Incremented for each file, starting at 0.
            uuid_str : string
                Set independently at each restart of the recorder.

        """

    def get_digital_metadata(self, channel_name, top_level_dir=None):
        """Return `DigitalMetadataReader` object for <channel_name>/metadata.

        By convention, metadata in Digital Metadata format is stored in the
        'metadata' directory in a particular channel directory. This method
        returns a reader object for accessing that metadata. If no such
        directory exists, an IOError is raised.

        Parameters
        ----------
        channel_name : string
            Name of channel, one of ``get_channels()``.

        top_level_dir : None | string
            If None, use *first* metadata path starting from the top-level
            directory list of the current DigitalRFReader object, in case there
            is more than one match. Otherwise, use the given path as the
            top-level directory.

        Returns
        -------
        DigitalMetadataReader
            Metadata reader object for the given channel.

        """

    def read_metadata(self, start_sample, end_sample, channel_name, method="ffill"):
        """Read Digital Metadata accompanying a Digital RF channel.

        By convention, metadata in Digital Metadata format is stored in the
        'metadata' directory in a particular channel directory. This function
        reads that metadata for a specified sample range by getting a
        DigitalMetadataReader and calling its `read` function.

        Parameters
        ----------
        start_sample : int
            Sample index for start of read, given in the number of samples
            since the epoch (time_since_epoch*sample_rate).

        end_sample : None | int
            Sample index for end of read (inclusive), given in the number of
            samples since the epoch (time_since_epoch*sample_rate). If None,
            use `end_sample` equal to `start_sample`.

        channel_name : string
            Name of channel to read from, one of ``get_channels()``.

        method : None | 'pad'/'ffill'
            If None, return only samples within the given range. If 'pad' or
            'ffill', the first sample no later than `start_sample` (if any)
            will also be included so that values are forward filled into the
            desired range.

        Returns
        -------
        OrderedDict
            The dictionary's keys are the sample index for each sample of
            metadata found between `start_sample` and `end_sample` (inclusive).
            Each value is a metadata sample given as a dictionary with column
            names as keys and numpy objects as leaf values.

        Notes
        -----
        For convenience, some pertinent metadata inherent to the Digital RF
        channel is added to the Digital Metadata, including:

            sample_rate_numerator : int
            sample_rate_denominator : int
            samples_per_second : np.longdouble

        """

    def get_continuous_blocks(self, start_sample, end_sample, channel_name):
        """Find continuous blocks of data between start and end samples.

        This is similar to `read`, except it returns the length of the blocks
        of continous data instead of the data itself.

        Parameters
        ----------
        start_sample : int
            Sample index for start of read, given in the number of samples
            since the epoch (time_since_epoch*sample_rate).

        end_sample : int
            Sample index for end of read (inclusive), given in the number of
            samples since the epoch (time_since_epoch*sample_rate).

        channel_name : string
            Name of channel to read from, one of ``get_channels()``.

        Returns
        -------
        OrderedDict
            The dictionary's keys are the start sample of each continuous block
            found between `start_sample` and `end_sample` (inclusive). Each
            value is the number of samples contained in that continous block
            of data.

        See Also
        --------
        read : Similar, except the data itself is returned.

        """

    def get_last_write(self, channel_name):
        """Return tuple of time and path of the last file written to a channel.

        Parameters
        ----------
        channel_name : string
            Name of channel, one of ``get_channels()``.

        Returns
        -------
        timestamp : float | None
            Modification time of the last file written. None if there is no
            data.

        path : string | None
            Full path of the last file written. None if there is no data.

        """

    def read_vector(self, start_sample, vector_length, channel_name, sub_channel=None):
        """Read a vector of data beginning at the given sample index.

        This method returns the vector of the data beginning at `start_sample`
        with length `vector_length` for the given channel and sub_channel(s).
        The vector is always cast to the smallest safe floating-point dtype no
        matter the original type of the data.

        This method calls `read` and converts the data appropriately. It will
        raise an IOError error if the returned vector would include any missing
        data.

        Parameters
        ----------
        start_sample : int
            Sample index for start of read, given in the number of samples
            since the epoch (time_since_epoch*sample_rate).

        vector_length : int
            Number of samples to read per subchannel.

        channel_name : string
            Name of channel to read from, one of ``get_channels()``.

        sub_channel : None | int, optional
            If None, the return array will contain all subchannels of data and
            be 2-d or 1-d depending on the number of subchannels. If an
            integer, the return array will be 1-d and contain the data of the
            subchannel given by that integer index.

        Returns
        -------
        array
            An array of floating-point dtype and shape (`vector_length`,) or
            (`vector_length`, N) where N is the number of subchannels.

        See Also
        --------
        read_vector_1d : Read data into a 1-d vector of floating-point type.
        read_vector_raw : Read data into a vector of HDF5-native type.
        read : Read continuous blocks of data between start and end samples.

        """

    def read_vector_raw(
        self, start_sample, vector_length, channel_name, sub_channel=None
    ):
        """Read a vector of data beginning at the given sample index.

        This method returns the vector of the data beginning at `start_sample`
        with length `vector_length` for the given channel. The data is returned
        in its HDF5-native type (e.g. complex integer-typed data has a
        stuctured dtype with 'r' and 'i' fields).

        This method calls `read` and converts the data appropriately. It will
        raise an IOError error if the returned vector would include any missing
        data.

        Parameters
        ----------
        start_sample : int
            Sample index for start of read, given in the number of samples
            since the epoch (time_since_epoch*sample_rate).

        vector_length : int
            Number of samples to read per subchannel.

        channel_name : string
            Name of channel to read from, one of ``get_channels()``.

        sub_channel : None | int, optional
            If None, the return array will contain all subchannels of data and
            be 2-d or 1-d depending on the number of subchannels. If an
            integer, the return array will be 1-d and contain the data of the
            subchannel given by that integer index.

        Returns
        -------
        array
            An array of shape (`vector_length`,) or (`vector_length`, N) where
            N is the number of subchannels.

        See Also
        --------
        read_vector : Read data into a vector of floating-point type.
        read_vector_1d : Read data into a 1-d vector of floating-point type.
        read : Read continuous blocks of data between start and end samples.

        """

    def read_vector_1d(self, start_sample, vector_length, channel_name, sub_channel=0):
        """Read a 1-d vector of data beginning at the given sample index.

        This method is identical to `read_vector`, except the default
        subchannel is 0 instead of None. As such, it always returns a 1-d
        vector of the smallest safe floating-point type.

        Parameters
        ----------
        start_sample : int
            Sample index for start of read, given in the number of samples
            since the epoch (time_since_epoch*sample_rate).

        vector_length : int
            Number of samples to read per subchannel.

        channel_name : string
            Name of channel to read from, one of ``get_channels()``.

        sub_channel : None | int, optional
            If None, the return array will contain all subchannels of data and
            be 2-d or 1-d depending on the number of subchannels. If an
            integer, the return array will be 1-d and contain the data of the
            subchannel given by that integer index.

        Returns
        -------
        array
            An array of floating-point dtype and shape (`vector_length`,).

        See Also
        --------
        read_vector : Read data into a vector of floating-point type.
        read_vector_raw : Read data into a vector of HDF5-native type.
        read : Read continuous blocks of data between start and end samples.

        """

    def read_vector_c81d(
        self, start_sample, vector_length, channel_name, sub_channel=0
    ):
        """Read a c8 1-d vector of data beginning at the given sample index.

        This method is similar to `read_vector`, except the default
        subchannel is 0 instead of None and the returned dtype is always
        complex64.

        Parameters
        ----------
        start_sample : int
            Sample index for start of read, given in the number of samples
            since the epoch (time_since_epoch*sample_rate).

        vector_length : int
            Number of samples to read per subchannel.

        channel_name : string
            Name of channel to read from, one of ``get_channels()``.

        sub_channel : None | int, optional
            If None, the return array will contain all subchannels of data and
            be 2-d or 1-d depending on the number of subchannels. If an
            integer, the return array will be 1-d and contain the data of the
            subchannel given by that integer index.

        Returns
        -------
        array
            An array of complex64 dtype and shape (`vector_length`,).

        See Also
        --------
        read_vector : Read data into a vector of floating-point type.
        read_vector_1d : Read data into a 1-d vector of floating-point type.
        read_vector_raw : Read data into a vector of HDF5-native type.
        read : Read continuous blocks of data between start and end samples.

        """

    @staticmethod
    def _get_file_list(
        sample0,
        sample1,
        samples_per_second,
        subdir_cadence_seconds,
        file_cadence_millisecs,
    ):
        """Get an ordered list of data file names that could contain data.

        This takes a first and last sample and generates the possible filenames
        spanning that time according to the subdirectory and file cadences.

        Parameters
        ----------
        sample0 : int
            Sample index for start of read, given in the number of samples
            since the epoch (time_since_epoch*sample_rate).

        sample1 : int
            Sample index for end of read (inclusive), given in the number of
            samples since the epoch (time_since_epoch*sample_rate).

        samples_per_second : np.longdouble
            Sample rate.

        subdir_cadence_secs : int
            Number of seconds of data found in one subdir. For example, 3600
            subdir_cadence_secs will be saved in each subdirectory.

        file_cadence_millisecs : int
            Number of milliseconds of data per file. Rule:
            (subdir_cadence_secs*1000 % file_cadence_millisecs) must equal 0.

        Returns
        -------
        list
            List of file paths that span the given time interval and conform
            to the subdirectory and file cadence naming scheme.

        """

    def _combine_blocks(self, cont_data_dict, len_only=False):
        """Order and combine data given as dictionary into continuous blocks.

        Parameters
        ----------
        cont_data_dict : dict
            Dictionary where keys are the start sample of a block of data and
            values are arrays of the data as found in a file. These blocks do
            not cross file boundaries and so may need to be combined in this
            method.

        len_only : bool
            If True, returned dictionary values are lengths. If False,
            values are the continuous arrays themselves.

        Returns
        -------
        OrderedDict
            The dictionary's keys are the start sample of each continuous block
            in ascending order. Each value is the array or length of continous
            data starting at the key's index.

        """

    def _get_channels_in_dir(self, top_level_dir):
        """Return a list of channel paths found in a top-level directory.

        A channel is any subdirectory with a drf_properties.h5 file.

        Parameters
        ----------
        top_level_dir : string
            Path of the top-level directory.

        Returns
        -------
        list
            A list of strings giving the channel paths found.
        """
</context_2_code_snippet>

<context_2_examples>
"""An example of reading Digital RF data in python.

Assumes the example Digital RF write script has already been run.

"""
from __future__ import absolute_import, division, print_function

import os
import tempfile

import digital_rf

datadir = os.path.join(tempfile.gettempdir(), "example_digital_rf")
try:
    dro = digital_rf.DigitalRFReader(datadir)
except ValueError:
    print("Please run the example write script before running this example.")
    raise

channels = dro.get_channels()
print("found channels: %s" % (str(channels)))

print("working on channel junk0")
start_index, end_index = dro.get_bounds("junk0")
print("get_bounds returned %i - %i" % (start_index, end_index))
cont_data_arr = dro.get_continuous_blocks(start_index, end_index, "junk0")
print(
    (
        "The following is a OrderedDict of all continuous block of data in"
        "(start_sample, length) format: %s"
    )
    % (str(cont_data_arr))
)

# read data - the first 3 reads of four should succeed, the fourth read
# will be beyond the available data
start_sample = list(cont_data_arr.keys())[0]
for i in range(4):
    try:
        result = dro.read_vector(start_sample, 200, "junk0")
        print(
            "read number %i got %i samples starting at sample %i"
            % (i, len(result), start_sample)
        )
        start_sample += 200
    except IOError:
        print("Read number %i went beyond existing data as expected" % (i))

# finally, get all the built in rf properties
rf_dict = dro.get_properties("junk0")
print(
    ("Here is the metadata built into drf_properties.h5 (valid for all data):" " %s")
    % str(rf_dict)
)
</context_2_examples>
</context_2>
</ReadMe.LLM>
\end{lstlisting}

\end{document}